\documentclass[journal]{IEEEtran}
\usepackage{amsmath}
\usepackage{graphicx}
\usepackage{cite}
\usepackage{amsfonts}
\usepackage{booktabs}
\usepackage{multirow}
\usepackage{flushend}
\usepackage{mathrsfs}
\usepackage{hyperref}
\usepackage{cleveref}

\usepackage{longtable, booktabs}
\usepackage{supertabular}
\usepackage{lscape}
\usepackage{eurosym}
\usepackage{verbatim}
\usepackage{soul, color, xcolor}
\usepackage{tikz}
\usetikzlibrary{arrows.meta,positioning}

\ifCLASSINFOpdf
\else
\fi

\begin{document}

\title{Secure Communications, Sensing, and Computing Towards Next-Generation Networks}

\author{
Ruiqi Liu,~\IEEEmembership{Senior Member},~IEEE,  Beixiong Zheng,~\IEEEmembership{Senior Member,~IEEE}, Jemin Lee,~\IEEEmembership{Senior Member,~IEEE}, Si-Hyeon Lee,~\IEEEmembership{Senior Member},~IEEE, Georges Kaddoum,~\IEEEmembership{Senior Member}, IEEE, \\Onur Günlü,~\IEEEmembership{Senior Member},~IEEE, and Deniz Gündüz,~\IEEEmembership{Fellow},~IEEE

\thanks{
R. Liu is with the State Key Laboratory of Mobile Network and Mobile Multimedia Technology, Shenzhen 518055, China and the Wireless and Computing Research Institute, ZTE Corporation, Shenzhen 518055, China (email: richie.leo@zte.com.cn).}

\thanks{
B. Zheng is with the School of Microelectronics, South China University of Technology, Guangzhou 511442, China (email: bxzheng@scut.edu.cn).
}

\thanks{
J. Lee is with the Department of Electrical and Electronic Engineering, Yonsei University, Seoul 03722, South Korea (email: jemin.lee@yonsei.ac.kr).}

\thanks{S.-H. Lee is with the School of Electrical Engineering, Korea Advanced Institute of Science and Technology (KAIST), Daejeon 34141, South Korea (email: sihyeon@kaist.ac.kr).}

\thanks{G. Kaddoum is with the Department of Electrical
Engineering, École de Technologie Supérieure, Université du Québec,
Montréal, QC H3C 1K3, Canada  (email: georges.kaddoum@etsmtl.ca).}

\thanks{O. G\"unl\"u is with the Lehrstuhl für Nachrichtentechnik, Technische Universit{\"a}t Dortmund, 44227 Dortmund, Germany and Information Theory and Security Laboratory (ITSL), Link{\"o}ping University, 581 83 Link{\"o}ping, Sweden (email: onur.guenlue@tu-dortmund.de).}

\thanks{D. Gündüz is with the Department of Electrical and Electronic Engineering, Imperial College London, London, United Kingdom (email: d.gunduz@imperial.ac.uk)}

\thanks{
The work of Beixiong Zheng was supported in part by the National Natural Science Foundation of China under Grant 62571193 and Grant 62331022, the Guangdong program under Grant 2023QN10X446 and Grant 2023ZT10X148, and the GJYC program of Guangzhou under Grant 2024D01J0079 and Grant 2024D03J0006.
The work of S.-H. Lee was supported in part by the National Research Foundation of Korea (NRF) grant (No. RS-2025-00561467) and in part by the Institute of Information \& Communications Technology Planning \& Evaluation (IITP) grant (No. RS-2024-00360387, Development of core security technology utilizing multimodal properties of wireless communication channels), all funded by the Korea government (MSIT). The work of O. Günlü was partially supported by the ZENITH Research and Leadership Career Development Fund under Grant ID23.01, EU COST Action 6G-PHYSEC, Swedish Foundation for Strategic Research (SSF) under Grant ID24-0087, and German Federal Ministry of Research, Technology and Space (BMFTR) 6GEM+ Transfer Hub under Grants 16KIS2412 and 16KISS005. The work of D. Gündüz was supported in part by UKRI under the projects AI-R (EP/X030806/1) and INFORMED-AI (EP/Y028732/1), and by the SNS JU project 6G-GOALS under the EU Horizon program (Grant Agreement No. 101139232).}

\thanks{Corresponding author: Beixiong Zheng.}
}

% make the title area
\maketitle
\begin{abstract}
Next-generation wireless networks are progressing beyond conventional connectivity to incorporate emerging sensing and computing capabilities. This convergence gives rise to integrated systems that enable not only uninterrupted communication, but also environmental awareness, intelligent decision-making, and novel applications that take advantage of these combined features. At the same time, this integration brings substantial security challenges. As computing, sensing, and communication become more tightly intertwined, the overall complexity of the system increases, creating new vulnerabilities and expanding the attack surface. The widespread deployment of data-heavy artificial intelligence applications further amplifies concerns regarding data security and privacy. 

This paper presents a comprehensive survey of security and privacy threats, along with potential countermeasures, in integrated wireless systems. We first review physical-layer security techniques for communication networks, and then investigate the security and privacy implications of semantic and pragmatic communications and their associated cross-layer design methodologies. For sensing functionalities, we pinpoint security and privacy risks at the levels of signal sources, propagation channels, and sensing targets, and summarize state-of-the-art defense strategies for each. The growing computational requirements of these applications drive the need for distributed computing over the network, which introduces additional risks such as data leakage, weak authentication, and multiple points of failure. We subsequently discuss secure coded computing approaches that can help overcome several of these challenges. Finally, we introduce unified security frameworks tailored to integrated communication–sensing–computing architectures, offering an end-to-end perspective on protecting future wireless systems. %The state-of-the-art and future directions are articulated in this paper, along with analysis on possible standardization impact.
\end{abstract}

\IEEEpeerreviewmaketitle

\section{Introduction}
\begin{figure}[t]
\centering
\includegraphics[width=0.75\linewidth]{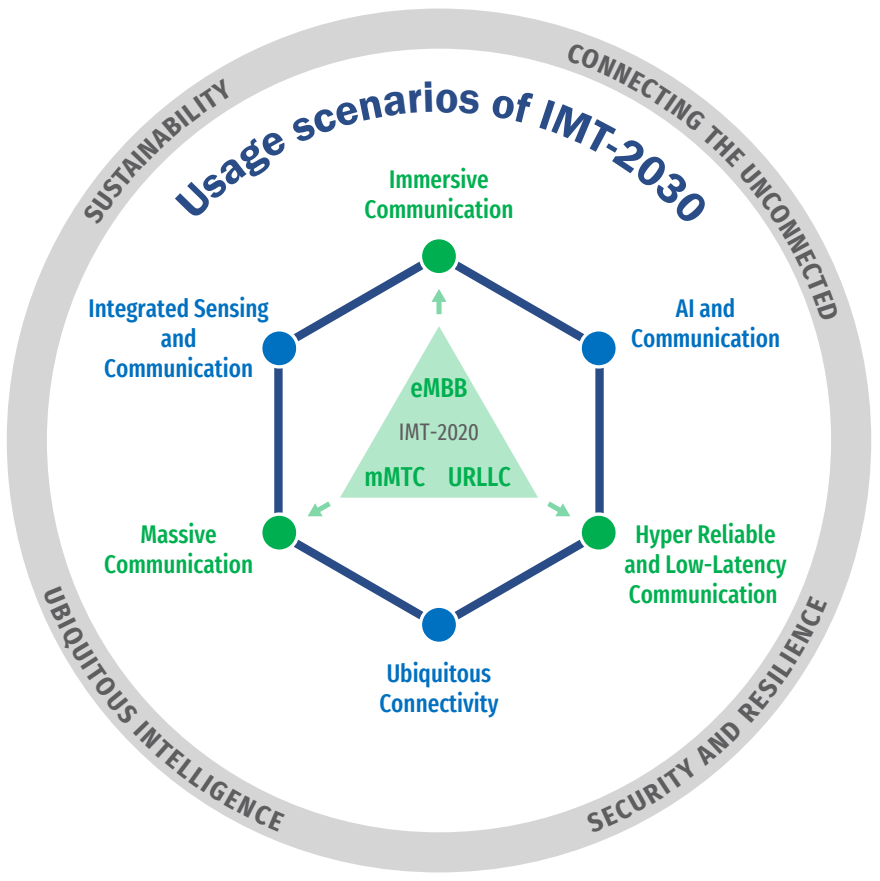}
\caption{A graphic illustration of the six usage scenarios and four overarching goals of 6G approved by the ITU-R. (Source: \cite{ITU_framework})}
\label{fig:usage}
\end{figure}

As the year 2030 approaches, the telecommunications industry and academia are working towards the next generation of mobile networks, namely, the International Mobile Telecommunications towards 2030 and beyond (IMT-2030), or 6G as more widely known, referring to the 6th generation of mobile communication networks. A recommendation is published by the International Telecommunication Union Radiocommunication Sector (ITU-R) as the IMT-2030 Framework \cite{ITU_framework}, providing general and technical guidelines for 6G designers and architects. A very high-level illustration of this recommendation is given in Fig. \ref{fig:usage}, defining six usage scenarios for 6G. Communication-centric usage scenarios such as immersive communication, massive communication and hyper reliable and low-latency communication (HRLLC) are still the main pillars of the system, while novel use cases are quickly emerging under integrated sensing and communication (ISAC) and artificial intelligence and communication (AIAC), and as immersive communication, that combines the two.

Security and privacy protection in communication networks is a classic research and engineering problem, which has been widely studied and standardized \cite{chen2012communication,33501,10183184}. The concept of security covers confidentiality, integrity, authenticity, and availability, and each aspect is protected by dedicated protocols or mechanisms. Towards 6G, further research efforts are needed to support stronger performance and ensure higher security at the same time \cite{10948395}. For instance, under the massive communication usage scenario, since connection density is expected to increase 5 to 10 times compared to the 5th generation (5G) \cite{10904090}, various improvements to current designs \cite{33861, 33851} will be needed to provide secure connectivity to a larger number of users. Similarly, for mission critical communications \cite{33880}, more advanced designs will be needed to guarantee higher reliability and security for HRLLC use cases, such as industrial control and remote healthcare. 
Lively discussions have been going on in standardization organizations such as the ITU and the 3rd generation partnership project (3GPP) around 6G security, studying potential candidate technologies including deep network slicing, quantum technologies, real-time adaptive security, distributed ledgers, and differential privacy \cite{9524814}.

As depicted in Fig. \ref{fig:usage}, a critical trend for 6G is the integration of artificial intelligence (AI) technologies into communication networks. This is reflected in the overarching concept of ubiquitous intelligence. Computing lies as the cornerstone in supporting AI functions and enabling intelligence, as data collection, training, and inference all involve intensive computing with stringent requirements. Integration of computing into communication systems will provide revolutionary capabilities, but also introduces many new security and privacy risks inherent in these computing systems into the communication networks, not only due to the large amount of data being collected and processed, but also to the many new capabilities the networks will acquire in understanding user behavior and patterns. In \cite{9404177}, researchers study and reveal that a variety of security and privacy risks can exist in computing systems, including data tempering and leakage, database intrusion, insecure communication among computing nodes and cyber attacks. Additionally, since massive amounts of data are generated and collected, data integrity, confidentiality, and availability can also suffer from insecure systems and cause potential risks \cite{ali2024information}. 

The integration of AI technologies into communication networks is also driving a fundamental paradigm shift toward semantic and goal-oriented communications \cite{Gunduz:JSAC:23, Calvanese:CB:21, Uysal:Network:22}. As AI applications become more widespread across networks, we can expect not just a dramatic surge in data traffic volume, but also a complete transformation in the types of data being transmitted and their associated requirements. Current networks are dominated by video content designed for human consumption, which requires transmitting substantial amounts of data to ensure high-quality reconstruction. However, communications intended for AI agents operate under fundamentally different principles. Rather than focusing on perfect reconstruction, the receiving AI system primarily needs to perform specific downstream tasks based on the underlying signal. This approach requires transmitting only the information that is relevant to these tasks, which represents the core concept of semantic and goal-oriented communication. On the other hand, avoiding high-resolution raw signal transmission can also enable inherent privacy protection, resulting in a fundamental trade-off between privacy and utility \cite{Calmon:Allerton:12, razeghi2023bottlenecks}: The objective is to enable legitimate receivers to perform their tasks with high accuracy while preventing the leakage of sensitive information.

Semantic communication shares deep connections with data compression. Traditional communication systems keep data compression and communication strictly separate, handling compression in the upper network layers. When maintained as separate processes, security for semantic communications can follow conventional approaches such as authentication and encryption. However, previous research has demonstrated that eliminating the boundary between semantic data compression and communication can deliver significant improvements \cite{bourtsoulatze2019deep, gunduz2024joint} in end-to-end signal quality and downstream task performance. This is achieved through neural network-based joint source-channel coding (JSCC) schemes, commonly known as DeepJSCC. Although DeepJSCC has attracted considerable research attention and its potential has been demonstrated through numerous prototype implementations \cite{Liu:VTC:22, Yoo:Access:23, Ding:ICCC:24, Abdi:INFOCOM:25}, it also creates new security challenges that must be addressed before its commercial adoption. The key advantage of JSCC schemes is the direct mapping of important input signal features to transmitted signals, resulting in an analog-like communication behavior. While this allows estimation of the input signal directly from the channel output, thus enabling signal reconstruction even in very poor channel conditions, it also makes the application of conventional digital encryption techniques challenging and increases the risk of data leakage.

ISAC is widely considered as a key usage scenario for 6G, including possible use cases of high accuracy localization, target detection on the ground and in the air \cite{10286534}, personal health monitoring \cite{ISAC_indoor_breath}, infrastructure monitoring, digital twins for smart cities, and intelligent transportation \cite{10440160}. Integration of communications and sensing can reduce the cost of deployment and operation for operators and, at the same time, provide new services to end users and vertical industries. As the integrated system can sense the surrounding physical world and detect and identify objects nearby, stakeholders raise new privacy and security concerns. In particular, ISAC systems can face challenges in protecting sensitive personal and operational data across multiple layers, including, but not limited to, transmitter privacy, target confidentiality, and environment privacy \cite{su2025integratingsensingcommunications6g}. In addition, illuminating targets with ISAC signals that carry information creates the opportunity for targets to eavesdrop on sensitive information \cite{SecureISAC,SecureISACLiU}. Non-cooperative and unauthorized sensors in the environment can also take advantage of ISAC signals emitted by authorized transmitters to sense the location, size, shape, and mobility of the target \cite{9417324}. 

Privacy is a critical issue to address before ISAC can become a reality \cite{su2023security}. As ISAC systems have the potential to sense and image the surrounding environment and all moving objects within their sensing range, proper methods to manage its applications and the usage of available data are essential. Moreover, in cooperative and even mono-static sensing modes, many network entities or functions will be involved in the sensing process, such as remote radio units (RRUs), baseband units (BBUs), core network, user equipment (UEs), and over-the-top applications \cite{10614082}, increasing the risk of leaking private data. Most existing communication security and privacy protection protocols rely on end-to-end encryption. These cannot be directly applied to ISAC systems as they involve unconnected objects. Thus, novel security protocols and even new secure sensing and data sharing technologies are needed to safely incorporate ISAC into 6G. 

As networks encounter new security challenges, architects of next-generation communication systems must consider security risks from the design phase to ensure the security and resilience of future networks. Researchers and engineers in both academia and industry are working to address these emerging security problems and develop advanced security protocols for system protection. This paper presents a comprehensive overview of these efforts: it not only overviews security and privacy risks associated with emerging communication paradigms that are expected to form key components of 6G networks, but also offers potential solutions and mitigation strategies. The new use cases and paradigms envisioned for 6G networks cannot be realized until the associated security and privacy concerns are adequately addressed, underscoring the critical importance of the research challenges examined in this work.

%In this paper, we start by introducing novel technology trends for the next generation networks, where communications, sensing and computing become three critical features. As a next step, effective  security protection measures for communications, sensing and computing are introduced, separately. 

The rest of the paper is organized as follows. Section \ref{Sec:secure_comm} presents effective security protection measures for physical layer communications as well as security protocols in higher layers.  
Section \ref{Sec:secure_sense} analyzes security aspects for sensing signal sources, sensing channels, and sensing targets. Section \ref{Sec:secure_compute} discusses secure coded computing, particularly focusing on integrated computing capabilities in 6G networks. In Section \ref{Sec:unified_design}, the unified security framework and designs are introduced for integrated communication, sensing, and computing systems where the three features can benefit from each other in terms of performance and security enhancement.
Finally, Section \ref{Sec:conclusion} concludes the paper.

\section{Secure Communications}\label{Sec:secure_comm}

\subsection{Physical Layer Secure Communications}\label{ss:phy_layer_security}

Physical layer connectivity forms the foundation of all communication systems. While security has traditionally been addressed at higher layers through digital encryption techniques, physical layer security aims to complement these cryptographic mechanisms by exploiting intrinsic characteristics of the wireless medium, including signal superposition, ambient noise, and channel randomness and reciprocity. Security in this context can encompass multiple objectives driven by the communication scenario. One fundamental goal is to prevent eavesdroppers from extracting any meaningful information about the content of legitimate communications. Another potential objective is to conceal the very existence of transmission, ensuring that a passive adversary cannot even detect that communication is taking place. These objectives correspond to the following two core aspects of physical layer security (PLS): the one related to the wiretap channel model and the other addressed through the covert communication framework. Beyond these passive threat models, physical layer techniques can also be extended to combat active attacks, such as impersonation or jamming, where adversaries aim to inject or disrupt signals, rather than merely observe them.
%to enhance the confidentiality, stealth, and availability of wireless transmissions. This approach capitalizes on the fact that wireless channels inherently introduce uncertainty and variability, which can be strategically exploited to safeguard sensitive information.

In this subsection, we focus on the two foundational scenarios involving passive adversaries: wiretap and covert communication. These represent distinct yet complementary objectives and highlight diverse capabilities of the physical layer in enhancing wireless security. In what follows, we explore both scenarios in further detail.

\subsubsection{Wiretap Channel}
The wiretap channel was first introduced by Wyner in his seminal work \cite{wyner1975wire}, which considered a three-terminal communication system consisting of a sender, an intended receiver, and an eavesdropper. The goal was to enable reliable communication to the intended receiver while ensuring that the eavesdropper gains asymptotically no information about the transmitted message. Wyner showed that when the legitimate channel is better than the eavesdropper’s channel in a stochastic sense, it is possible to achieve a strictly positive secrecy rate. This model was extended by Csiszár and Körner to the non-degraded case \cite{1055892}. %They characterized the secrecy capacity that involves auxiliary random variables in the single-letter expressions, laying the theoretical foundation for many follow-up works on secrecy in multi-terminal channels.

Since then, the basic wiretap framework has been extended in several important directions. One prominent line of work considered scenarios where some or all terminals are equipped with multiple antennas. In particular, in \cite{5485016, 5605343}, Khisti and Wornell established a computable characterization of the secrecy capacity of the multi-input, single-output, multi-eavesdropper (MISOME) channel and the multi-input, multi-output, multi-eavesdropper (MIMOME) channel, respectively. In the high signal-to-noise (SNR) regime, a simple masked beamforming scheme was demonstrated to be near optimal for the MISOME channel, while it could be arbitrarily far from capacity for the MIMOME channel. 
Another important extension involved communication over fading channels, which are more representative of practical wireless environments \cite{1523727,4529264,4626059}. In the case of slow fading channels, several works proposed an outage-based secrecy formulation \cite{1523727,4529264}, where the goal was to guarantee secure communication during the time instances when the main channel was stronger than the eavesdropper’s channel. By contrast, the work of Gopala, Lai, and El Gamal \cite{4626059} focused on delay-tolerant scenarios and adopted an ergodic formulation of secrecy capacity by coding over many fading blocks. Remarkably, their results revealed that it is possible to achieve a strictly positive perfectly secure rate even when the eavesdropper’s channel is, on average, better than the main channel, provided that appropriate power and rate adaptation strategies are employed.

While the secrecy capacity is relatively well understood in many three-terminal settings, it remains largely unresolved for multi-user network scenarios. For example, in multiple access channels, relay networks, and interference channels with secrecy constraints, the exact secrecy rate region is known only for a few special cases \cite{4675718,4797670,7105936,7302550}. Given this difficulty, a more tractable and fundamental metric often considered is the secure degrees of freedom (SDoF), which captures the asymptotic scaling of the secrecy rate in the high SNR regime. Notable contributions in this area include the works of \cite{6702446, 6783700, 7056440, 7593236, 7727973, 8443409}, which illustrated how spatial, temporal, and cooperative resources can be harnessed to ensure secure communication.

In practical communication scenarios with a large number of users and adopting various emerging technologies, past research gradually shifted from characterizing the optimal secrecy capacity region or SDoF to exploring the secrecy benefits and challenges introduced by new physical-layer techniques and architectures. In particular, reconfigurable intelligent surfaces (RIS) were reported to enhance secrecy rates with an increase of the number of reflecting elements \cite{9439833}, making them a promising tool for secure communications.
Furthermore, flexible antenna architectures, such as rotatable antenna (RA), emerged as a promising technology to enhance spatial degrees of freedom by dynamically reconfiguring antenna orientation/shape/position \cite{zheng2025rotatable,zheng2025rotatablemag,xiong2025intelligent,zheng2026Tutorial}.  
In secure communication systems, RA enables the control of the antenna's boresight direction that focuses energy towards legitimate users while suppressing signal leakage to eavesdroppers, significantly enhancing secrecy performance \cite{dai2025rotatable,zheng2026Tutorial}.  
Non-terrestrial networks (NTNs), including satellite systems and UAV-enabled communication platforms, introduce both promising opportunities and significant challenges from a PLS perspective \cite{8883128, 10636237, 9781300}. 
Regarding opportunities, NTNs offer highly controllable network elements, such as UAVs with flexible trajectories and altitudes, which can be dynamically optimized to improve the legitimate user’s channel while degrading the eavesdropper’s channel \cite{8883128, 10636237}. Their line-of-sight (LoS) dominant links and mobility patterns can be exploited for channel-aware beamforming, artificial noise design, and trajectory optimization, thus enabling more precise physical-layer strategies for enhancing secrecy. On the other hand, the authors of  \cite{9781300} identified several critical challenges to applying PLS in satellite-terrestrial integrated systems, such as high correlation between legitimate and eavesdropper channels due to broad satellite beams, co-channel interference from spectrum sharing, and unreliable CSI caused by long propagation delays and hardware impairments. To overcome these issues, the authors discussed solutions such as dual-beam dual-frequency schemes, secure beamforming, and cooperative relay techniques tailored for hybrid satellite-terrestrial environments.
Moreover, the impact on secrecy performance of next-generation communication technologies--such as massive multiple input multiple output (MIMO), millimeter-wave (mmWave) communications, non-orthogonal multiple access (NOMA), and full-duplex transmission--has been the subject of growing research interest \cite{7572045, 8245831, 7222458}. A comprehensive overview of these technologies in the context of PLS can be found in \cite{8335290}.

\subsubsection{Covert Communication}
Recent years have seen growing interest in reliable communication under the constraint that an adversary--frequently referred to as a warden--remains unaware of the presence of communication. This problem, studied under the names covert communication, low probability of detection (LPD), or undetectable communication, aims to characterize the maximum achievable throughput while keeping the adversary’s observations statistically indistinguishable from those under no transmission. Bash et al. \cite{6584948} showed that, under perfect channel knowledge, the number of covert bits that can be reliably transmitted via $n$ uses of an AWGN channel scales no faster than $\sqrt{n}$--a result known as the square root law (SRL).
More recently, the optimal constant in front of the $\sqrt{n}$ scaling for point-to-point covert communication was characterized \cite{7447769,7407378}.

The study of covert communication has been extended to multi-user communication settings involving multiple legitimate users and/or adversaries.  The fundamental limit of covert communication was studied for canonical network models such as multiple-access channels \cite{8768393,9828479}, broadcast channels \cite{8513890}, and interference channels \cite{9153017}. Notably, for broadcast and interference channels, the capacity region without covertness constraints remains largely unknown. Nevertheless, in the case of broadcast channels, a simple time-division scheme was shown to be optimal for a non-negligible class of channels under covert constraints \cite{8513890}. Interestingly, this class is not necessarily a subset or superset of the more capable class for which the capacity is known in the absence of covert constraints. For interference channels, it was established that treating interference as noise can achieve the covert communication capacity \cite{9153017}. The optimality of such simple strategies in both cases stems from the highly sparse nature of signaling required under covertness constraints. In large networks with many users and wardens, the scaling laws of covert capacity have been investigated. In static networks without node mobility, \cite{9146783} reported that optimal scaling can be achieved by introducing preservation regions around each warden and applying a modified version of the hierarchical cooperation scheme \cite{4305409}, which is known to be optimal without covertness constraints. When node mobility is allowed, the achievable covert throughput significantly improves as compared to the static case \cite{9296244}, thus highlighting the beneficial role of mobility in large-scale covert communications.

Several further studies explored the impact of channel uncertainty on covert communication. In particular, the works \cite{6970786, 7084182, 7964713} considered scenarios, where the channel’s noise level (or crossover probability) is random, remains fixed throughout the communication, and is unknown to the warden. Under this assumption, it becomes significantly more difficult for the warden to distinguish between noise and actual signals. As a result, positive covert communication rates can be achieved for certain channel models that would otherwise only support square-root scaling in the absence of noise uncertainty.
Furthermore, in \cite{8322303}, the authors studied covert communication in the presence of channel uncertainty represented by random channel states, demonstrating that nonzero covert rates are achievable when the transmitter has access to the channel state information (CSI), either in a causal or non-causal manner. The insight that channel uncertainty can enhance covert communication laid down the foundation for further research on incorporating jamming signals \cite{9381893,10188892},  transmit power variation \cite{9146329}, and location variation \cite{8764452}, to obscure transmission and improve covertness.

Similarly to wiretap setup, covert communication has  been actively extended to emerging communication scenarios \cite{10090449}. In particular, RIS showed promise in enhancing covertness by controlling the wireless environment to strengthen the legitimate link while minimizing signal leakage to the warden \cite{9438645, 9527330}. However, because of limited CSI and the passive nature of the surface, optimizing RIS configurations under covertness constraints remains challenging. NTNs offer mobility and altitude diversity that can be used to dynamically adjust transmission geometry and evade detection \cite{9456902,9536435,10520820}. Yet, the strong LoS  channels common in NTNs may inadvertently assist the warden, especially in wide coverage areas. NOMA also presents new opportunities, where covert signals can be embedded within superposed transmissions to mask their presence \cite{9146329,9870641}. These developments highlight the growing interest in applying covert communication principles to next-generation wireless systems, while also revealing new technical challenges that diverge from conventional settings.

\subsection{Physical Layer Key Generation}
%{\color{red}[Jemin]}
Cryptographic security solutions, such as encryption and decryption, rely on secret keys that need to be securely distributed and managed for renewal and revocation. However, key distribution and management often impose significant overhead on communication networks, and carry inherent risks of key leakage. To address these challenges, physical layer key generation (PLKG) has emerged as a promising alternative, enabling the generation of shared secret keys from the unique and random characteristics of wireless channels.

The performance of PLKG is commonly evaluated using two key metrics: the key generation rate (KGR), which denotes the number of secret bits generated per second, and the key disagreement rate (KDR), which represents the proportion of mismatched bits between the keys generated by legitimate users. A high KGR is essential for supporting real-time cryptographic protocols that demand long keys or frequent key refreshes. Meanwhile, a low KDR is critical for successful key agreement; otherwise, the protocol fails to establish a consistent shared key \cite{24_IoT_LSTM}, \cite{23_IoT_FFAencoder}.

\subsubsection{Key Generation Procedures}
There are generally five steps in the key generation as follows: a) channel probing, b) pre-processing, c) quantization and key generation, d) information reconciliation, and e) privacy amplification.

\begin{itemize}
    \item \textbf{Channel Probing:}
    Channel probing refers to the exchange of pilot signals between legitimate users to estimate the wireless channels between them. Owing to the property of channel reciprocity within the coherence time, both users can obtain similar channel measurements, which serve as the basis for secret key generation.
    Although theoretical analyses assume the channel independency with the separation at least half a wavelength, in practice, the broadcast nature of pilot signals may cause information leakage. This is especially concerning when an attacker can be located near one of the legitimate users, potentially allowing partial channel observation.\\
    {To mitigate this leakage, the protective token-based PLKG scheme\cite{25_TIFS_PT} can be employed. This method encapsulates the pilot signals with protective tokens, preventing eavesdroppers from accessing the concealed pilot information. In addition, random pilot activation 
    schemes \cite{24_TVT_RPA, 24_TWC_RPA}, which activate only a subset of pilot tones, can be used to further reduce both information leakage and pilot overhead.} 
    %The legitimate receiver can then reconstruct the full channel information through interpolation.}

    \item \textbf{Pre-processing on Channel Information:}
    In real-world environments, perfect channel reciprocity is difficult to achieve due to user mobility, hardware impairments, background noise, and the Doppler effect. This non-reciprocity can lead to a high KDR between legitimate users. The goal of the pre-processing step is to refine the measured channel information before it is used for key generation, reducing the KDR.\\
    {To enhance the reciprocity of the measured channel features, deep learning-based techniques have been explored. For example, a hybrid model that combines a long short-term memory (LSTM) network and a Kalman filter has been proposed to predict and smooth time-varying channel measurements \cite{24_IoT_LSTM}. The feature fusion autoencoder (FFAEncoder) has been used to extract and fuse amplitude and phase components separately, effectively mitigating noise interference \cite{23_IoT_FFAencoder}. Similarly, the bidirectional convergent feature learning (BCFL) convolutional neural network (CNN) jointly processes the CSI of both legitimate users to extract shared features, maximizing reciprocity \cite{23_IoT_BCFL}.}

    \item \textbf{Quantization and Key Generation}
    This step converts continuous-valued analog channel features into binary bit strings, which serve as the secret key. A quantizer is first applied to map the continuous channel measurements to discrete values. Typically, a double-threshold quantization scheme is used, where two thresholds, $\nu_1$ and $\nu_2$, are defined. A sample value is mapped to bit ‘1’ if it exceeds $\nu_1$, or to bit ‘0’ if it is below $\nu_2$, while samples falling between thresholds are discarded. To improve key generation efficiency, the discarded samples can be reused by feeding them into the quantizer in subsequent rounds \cite{23_IoT_BCFL}.

    \item \textbf{Information Reconciliation and Privacy Amplification}
   \emph{Information reconciliation} aims to resolve mismatches in the initial key sequences generated by legitimate users after the quantization step. Commonly used algorithms for this consistency negotiation include Cascade, Bose–Chaudhuri–Hocquenghem (BCH) codes \cite{23_ComSoc_UAV}, and other error correction coding techniques. Additionally, fuzzy extractors can also be employed to correct key mismatches by producing identical random outputs from two inputs that are nearly identical, such as noisy or imperfect versions of the same key \cite{20_TIT_Fuzzy}, \cite{24_IoT_LSTM}.
   \emph{Privacy amplification} aims to eliminate any potential information leakage that may have occurred during the public exchange process in the information reconciliation phase. This is typically achieved by applying a strong one-way hash function to the reconciled key sequence, which compresses the key into a shorter but highly random output. The resulting key is information-theoretically secure, ensuring that any partial knowledge an eavesdropper may have is rendered useless. For instance, cryptographic hash functions from the SHA-2 family, such as SHA-256, are commonly employed for this purpose \cite{24_IoT_LSTM}, \cite{24_TWC_RPA}.
    
\end{itemize}

\subsubsection{Physical Sources for Key Generation}

Various physical-layer characteristics of wireless channels can serve as entropy sources for secret key generation.
First, CSI refers to a set of complex-valued parameters that describe how a signal propagates from a transmitter to a receiver. The CSI captures fine-grained amplitude and phase variations across different frequencies or time instances. Its high dimensionality allows for a large amount of data to be extracted from a single measurement, while its sensitivity to small environmental changes ensures high randomness, which is desirable for security purposes \cite{23_IoT_FFAencoder}.

The CSI can be represented either in the frequency domain as the channel frequency response (CFR) \cite{23_ComSoc_UAV} or in the time domain as the channel impulse response (CIR)\cite{23_Access_MIMO}, which reflects the multipath structure of the channel. While CIR provides rich temporal detail, its effectiveness can be limited in highly dynamic scenarios. 
%Nevertheless, various techniques have been proposed to exploit CIR in MIMO-OFDM systems \cite{23_Access_MIMO}.
In contrast, the received signal strength (RSS) is a coarse-grained alternative, offering a single scalar value representing average signal power. Although RSS is universally available and easy to measure, its low entropy limits the KGR. As such, RSS is mainly suitable for highly mobile or resource-constrained environments, such as vehicular ad hoc networks (VANETs), where complex channel estimation is infeasible \cite{24_IoT_LSTM}.
The propagation delay can also be utilized as a source of randomness in specialized scenarios such as underwater acoustic networks \cite{23_TIFS_UWAN}. 
%For instance, \cite{23_TIFS_UWAN} proposes a key generation method based on the propagation delay characteristics of underwater acoustic networks (UWANs).

\subsubsection{Challenges of PLKG} 
The practical deployment of PLKG faces several challenges. A primary issue is that many existing methods are evaluated under idealized or unrealistic assumptions, which often do not hold in dynamic real-world environments \cite{24_TVT_FDD}. In addition, the physical characteristics of the wireless environment can significantly limit the effectiveness of PLKG. For example, scenarios with limited multipath propagation,  such as air-to-air (A2A) UAV communication links \cite{23_ComSoc_UAV} or static indoor settings, may fail to provide sufficient channel randomness for secure and reliable key generation.

\subsection{Physical Layer Authentication}
%Authentication is a crucial concern in wireless communication due to the increasing security threats associated with its rapid development. 
%
The conventional authentication is achieved via upper-layer authentication, which uses cryptography-based algorithms. Such authentication generally requires high communication and computation overhead, which is difficult to be used for massive networks. 
Physical layer authentication (PLA) is developed to authenticate users based on the physical characteristics of its signal, rather than relying on digital passwords or cryptographic keys.  

The performance of the PLA is generally evaluated by the false alarm rate (FAR) (i.e., false positives) and the miss detection rate (MDR) (i.e., false negatives). The trade-off between FAR and MDR is commonly illustrated using the receiver operating characteristic (ROC) curve, where a larger area under the curve (AUC) indicates better authentication performance.
As a more sophisticated performance evaluation, a confusion matrix is also used,  
comprising true positives, true negatives, false positives, and false negatives \cite{23IOT_RuiMeng}. 
%In addition to classification accuracy, the decision time is also considered as an important performance metric, especially for real-time applications\cite{19JSAC_Forssell}.

%Existing PLA schemes can be categorized into two categories: passive and active schemes. In the passive schemes, a receiver authenticates the transmitter based on the physical-layer features of the received signals. In the active schemes, a transmitter generates a tag based on a secret key and embeds it into a transmitting signal. 
\subsubsection{Authentication Procedures}
There are generally three steps on the PLA, which consists of three stages as follows.

\begin{itemize}
    \item \textbf{Training}: 
    This stage is initiated after an upper-layer authentication confirms the user’s identity. The authenticated user is then allowed to transmit predefined training signals, enabling the receiver to extract physical-layer features such as CSI or RF fingerprints. These features are used to build a reference profile, which is stored in the white-list for future authentication.
    \item \textbf{Data Transmission}: An unauthenticated user transmits a signal, from which the receiver extracts physical-layer features and compares them against the reference profiles stored in the white-list.
    \item \textbf{Authentication}: Based on this comparison, the user is classified as legitimate or unauthorized. This authentication process can be implemented using two main approaches: 1) statistical hypothesis testing and 2) machine learning-based authentication.
\end{itemize}

\emph{Statistical hypothesis testing} is a threshold-based method that makes decisions by evaluating a similarity or distance metric between the extracted feature values and reference data\cite{16TWC_Liu}. The decision threshold can be determined either analytically to meet a target authentication performance (e.g., the false alarm rate) or empirically through validation experiments.

The \emph{machine learning-based authentication} has been used with the development of machine learning (ML), specifically, the classification algorithms such as k-nearest neighbor (k-NN), support vector machine (SVM), k-means clustering (K-means), and one-class support vector machine (OC-SVM). For instance, the SVM can be used to identify users based on their radio frequency-distinct native attribute fingerprints\cite{21IoT_Reising}. 
The deep learning algorithms have also been incorporated into authentication schemes. For instance, neural networks \cite{19ICAICA_Liao} and generative adversarial networks (GANs) \cite{20ICSPCS_Germain} have been proposed for user authentication, demonstrating improved performance in extracting and learning complex physical-layer features.

\subsubsection{Physical Sources for Authentication}
The physical sources of authentication can be divided into three categories: 1) device-based features, 2) channel-based features, and 3) hybrid features. 

\begin{itemize}
    \item \textbf{Device-based Features}: 
    The device-based features are derived from unique hardware-induced imperfections of transmitters \cite{12ICC_Hou}. This feature includes  transient amplitude, carrier frequency offset, phase noise, and I/Q imbalance. 
    %
    %\cite{12ICC} proposed a CFO-based authentication method in a time-invariant OFDM system. Hypothesis testing is applied to identify legitimate transmitters based on the CFO estimated from received signals. %Thresholds are determined based on SNR and the method demonstrates robust performance under multipath fading channels. 
%
    \item \textbf{Channel-based features}:
     The channel-based features can be categorized into 1) statistical channel information (e.g., RSS \cite{19Access_Qiu}) and 2) instantaneous channel information (e.g., CIR and CFR \cite{08ICC_Xiao}). Among them, the RSS is the most widely used attribute due to its stability and ease of acquisition. 
    \item \textbf{Hybrid Features}:
    Two or more features can be used jointly in authentication to enhance reliability and robustness\cite{20TC_Zhang}. For instance, for authentication, a fingerprint of a transmitter can be obtained as a weighted combination of multiple device-based features \cite{19IoT_Peng} or both device- and channel-based features \cite{20TC_Zhang}.   
\end{itemize}

\subsubsection{Challenges of PLA}

%The PLA can provide valuable advantages over traditional cryptography-based authentication, and its performance can be further enhanced by incorporating ML techniques. However, several challenges remain.
%
One of the main challenges is the feature selection. Many physical-layer features, especially those derived from wireless channels, are highly sensitive to environmental dynamics such as mobility, obstacles, or interference. Ensuring consistent authentication performance under varying conditions is a significant difficulty.
Moreover, the PLA is typically performed without prior knowledge of attacker characteristics. Hence, the authentication framework must be designed to generalize well to a wide range of potential attack strategies, including unforeseen ones.
In particular, the emergence of digital twin technology introduces new security concerns in this aspect. By replicating the real-world wireless environment, digital twins may enable attackers to better observe and learn the statistical properties of the legitimate channel. This, in turn, can enhance the attacker’s ability to launch effective spoofing attacks.

\subsection{Physical Layer Quantum Key Distribution}
Among many physical layer key distribution methods, quantum key distribution (QKD) has the unique property since its security is not based on computational assumptions but on the laws of physics and quantum mechanics \cite{science.283.5410.2050}. Attempting to eavesdrop the QKD channel would inevitably disturb the quantum states of photons and can be detected by the legitimate receiver. Thanks to this property, QKD is studied by both academia and industry for a variety of scenarios and use cases. QKD was initially designed to be implemented in wired optical networks using single photons \cite{Hughes01022000}, later on extended into wireless optical networks and even radio-frequency wireless networks through continuous-variable QKD \cite{9684555}.

There are several standardization organizations that work on QKD for communication systems, including the European Telecommunications Standards Institute (ETSI) and the International Telecommunication Union (ITU) Telecommunication Standardization Sector (ITU-T). 
In ETSI, an industry specification group (ISG) was established to address research and pre-standard topics related to QKD. The ISG meets regularly to develop group reports (GRs) and group specifications (GSs) describing quantum cryptography for information and communication networks. GRs capture informative work, such as technical overview on specific topics while GSs capture normative work that will have stronger impact on system implementations. In ETSI ISG QKD, the work items focus on QKD for wired optical networks.
The primary focus of ITU-T SG 13 is future network, and QKD falls into this scope as a novel technology for 5G and beyond. The deliverables published by ITU-T SG 13 focus more on technology overview and overall network architecture design. There is another group, ITU-T SG 17, which focuses particularly on security, also studying quantum related security enhancement. Work items in SG 17 focus specifically on framework, architecture, requirements, authentication, authorization and management on security enabled by QKD.
Some work items are summarized in Table \ref{table:QKD}, while more comprehensive review on standardization for QKD can be found in the literature \cite{qtc2.12044,qtc2.12069}.

\begin{table*}[thb]
  \begin{center}
	\caption{Summary of selected deliverable on QKD standardization}
	\label{table:QKD}
	\begin{tabular}{|p{55pt}|p{50pt}|p{130pt}|p{175pt}|}
		\hline
	Standardization group & Identifier & Title & Scope \\
     \hline
       ETSI ISG QKD & GS QKD 011 &  Component characterization: characterizing optical components for QKD systems  & Specifying and providing procedures for the characterization of optical components in QKD systems  \\
       \hline
    ETSI ISG QKD & GS QKD 012 & Device and Communication Channel Parameters for QKD Deployment  & Describing the main communication resources in QKD systems and providing a possible architecture for QKD deployment over an optical network infrastructure  \\
        \hline
	ETSI ISG QKD & GR QKD 017 &  Network architectures &  Reviewing the variety of architectures that have been proposed for QKD networking  \\
		\hline
        ITU-T SG 13	& Y.3800 & Overview on networks supporting quantum key distribution & Providing conceptual layer structures of QKDN, and defining the basic of different layers in QKDN \\
        	\hline
	ITU-T SG 13	& Y.3802 & Quantum Key Distribution networks - Functional architecture & Defining a functional architecture model of QKDN, specifying detailed functional elements and reference points, architectural configurations and basic operational procedures of QKDN \\
    	\hline
	ITU-T SG 13	& Y.3804 & Quantum key distribution networks - Control and management & Specifying control and management functions and procedures of QKDN control, management, and orchestration distribution \\
            \hline
	ITU-T SG 17	& X.1710 & Security framework for quantum key distribution networks & Specifying a simplified framework including security requirements and measures to combat security threats for QKDN \\
		\hline
	ITU-T SG 17	&	X.1712 & Security requirements and measures for QKD networks - key management &  Specifying the security threats, security requirements and security measures of key management for QKDN \\
		\hline
       ITU-T SG 17	&  X.1713 & Security requirements for the protection of quantum key distribution nodes & Identifying security threats, providing security requirements for QKD nodes and providing specific techniques to meet the requirements \\
        \hline
	ITU-T SG 17	&	X.1714 & Key combination and confidential key supply for quantum key distribution network & Describing key combination methods for QKDN and specifying security requirements for both key combination and key supply \\
        \hline
        ITU-T SG 17	& X.1716 & Authentication and authorization in quantum
key distribution network & Studying IDs and their management, public key certification supported by the Public Key Infrastructure (PKI), and authentication and authorization in QKDN \\
	    \hline
	   ITU-T SG 17	& X.1717 & Security requirements and measures for quantum key distribution networks (QKDN) - control and management &  Specifying use cases, security threats in the context of quantum computing, security requirements and security measures for controllers and managers of QKDN \\
        \hline
    
		\hline
	\end{tabular}
  \end{center}
\end{table*}

%%%%%%%%%%%%%%%%%%%%%%%%%%%%%%%%%%%%%%%%%%%%%%%%%%%%%%%%%%%%%%%%%%%%%%%%%%
%%%%%%%%%%%%%%%%%%%%%%%%%%%%%%%%%%%%%%%%%%%%%%%%%%%%%%%%%%%%%%%%%%%%%%%%%%

\subsection{Security and Privacy in Semantic Communications}\label{ss:SP_Semantics}

Semantic and goal-oriented communication focuses on transmitting only information relevant to the receiver's downstream tasks \cite{Gunduz:JSAC:23, Lu:CST:24}, typically involving cross-layer design. Consider, as an example, transmitting a photo where the receiver only needs to identify the person shown - the background becomes irrelevant and need not be transmitted. Under this broad definition, any lossy compression scheme constitutes semantic communication, as it removes parts of the source signal that do not affect the prescribed distortion measure, without impacting the end-to-end performance. For example, lossy image compression removes imperceptible high-frequency components without degrading perceived image quality.

However, classical lossy compression schemes are generally considered in the context of source reconstruction; that is, the receiver aims to recover the input signal with maximum fidelity. Mathematically, classical compression problems assume that source and reconstruction share the same alphabet, while semantic communications allow arbitrary reconstruction alphabets. Many emerging applications involving edge intelligence tasks require receivers to perform inference rather than reconstruction. Using the image example, receivers might identify objects, count people or vehicles, retrieve similar images from local datasets, or perform image segmentation. These problems can be formalized as lossy compression with general distortion measures where reconstruction and source alphabets differ—for image classification, the input alphabet represents pixel tensors while the reconstruction alphabet contains class labels.

This paradigm shift toward semantic communication offers significant improvements in bandwidth efficiency, latency reduction, and system performance. However, it introduces new security and privacy challenges that traditional cryptographic approaches cannot fully address. Semantic communications extract and transmit semantic features rather than raw data, often using deep neural networks (DNNs) to encode task-relevant information while discarding irrelevant details. Integration of AI and ML components, semantic feature extraction processes, and context-aware transmission strategies create novel attack surfaces and vulnerability vectors, which require new security frameworks.

\subsubsection{Privacy Challenges and Solutions}\label{ss:PUT}

Traditional communication systems primarily address security threats from passive eavesdroppers or active adversaries (see Subsection \ref{ss:phy_layer_security} for further details). When semantic communication operates at the application layer through feature extraction and lossy compression, the security of compressed features against eavesdroppers can be addressed using conventional approaches. However, privacy concerns arise when considering the content of the source signal itself. The source may contain sensitive information that the transmitter prefers not to share even with legitimate receivers of securely transmitted data. Even when transmitted features remain protected, they may inadvertently reveal sensitive information. For example, facial recognition features could expose identity information without reconstructing the original image. Therefore, private semantic communication requires identifying the most relevant features for a given receiver task while simultaneously ensuring the privacy of sensitive aspects. Since these objectives often conflict, system designers must operate at a desired point on the associated privacy-utility trade-off (PUT). 

Privacy-preserving mechanisms add noise or randomize data to prevent statistical inferences before sharing with third parties. These mechanisms are evaluated using privacy metrics that account for adversary types and available data sources. Privacy-preserving approaches are classified as either `prior-independent' (i.e., making minimal assumptions about data distributions and adversary capabilities) or `prior-dependent' (i.e., using knowledge of private data probability distributions and adversary abilities) \cite{razeghi2023bottlenecks}.

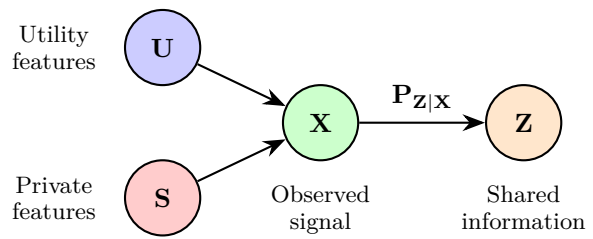
\begin{figure}[!t]
    \centering

    \begin{tikzpicture}[
        mynode/.style={draw, circle, minimum size=1.0cm, thick},
        arrow/.style={-{Stealth[length=3mm]}, thick},
        label/.style={font=\small, align=center}
    ]
  
    % Define nodes
    \node[mynode, fill=blue!20] (utility) at (-2.1,1) {$\mathbf{U}$};
    \node[label, left=0.25cm of utility] {Utility\\features};
    
    \node[mynode, fill=red!20] (private) at (-2.1,-1) {$\mathbf{S}$};
    \node[label, left=0.25cm of private] {Private\\features};
    
    \node[mynode, fill=green!20] (observed) at (0,0) {$\mathbf{X}$};
    \node[label, below=0.2cm of observed] {Observed \\ signal};
    
    \node[mynode, fill=orange!20] (compressed) at (2.7,0) {$\mathbf{Z}$};
    \node[label, below=0.2cm of compressed] {Shared \\ information};
    
    % Draw arrows
    \draw[arrow] (utility) -- (observed);
    \draw[arrow] (private) -- (observed);
    \draw[arrow] (observed) -- (compressed) node[midway, above] {$\mathbf{P_{Z|X}}$};
    
    \end{tikzpicture}

    \caption{Illustration of the complexity-leakage-utility bottleneck (CLUB).}
    \label{fig:PUT}
\end{figure}

Data anonymization approaches include well-known schemes such as $k$-anonymity \cite{sweeney2002k}, $\ell$-diversity \cite{machanavajjhala2006diversity}, $t$-closeness \cite{li2007t}, differential privacy (DP) \cite{dwork2006calibrating}, and pufferfish privacy \cite{kifer2012rigorous}, all of which rely on data perturbation. DP, the most popular prior-independent privacy notion, is characterized by distinguishability of `neighboring databases. Alternatively, information-theoretic (IT) privacy studies mechanisms and metrics that preserve privacy when statistical properties of data can be estimated, or partially known. IT privacy \cite{reed1973information, yamamoto1983source, makhdoumi2014information,  sreekumar2019optimal, rassouli2021perfect} analyzes PUT using IT metrics that quantify adversarial information gain about private features from disclosed data. These metrics, often formulated as divergences between probability distributions (e.g., f-divergences, Renyi divergence, or total variation distance \cite{Rassouli:TIFS:20}), balance useful information extraction with privacy preservation. The IT privacy framework draws inspiration from Shannon's information-theoretic secrecy \cite{shannon1949communication} and Reed \cite{reed1973information} and Yamamoto's \cite{yamamoto1983source} lossy source coding treatment of security and privacy. It has been applied to distributed lossless compression in \cite{Prabhakaran:ITW:2007, Gunduz:ISIT:08}. In \cite{makhdoumi2014information}, it was formalized as the privacy funnel framework, which can be considered as a dual of the information bottleneck principle \cite{tishby2000information}. These two frameworks were combined and extended in \cite{razeghi2023bottlenecks} by treating both utility and private signals as latent variables correlated with the observed signal, resulting in the three-way complexity-leakage-utility bottleneck (CLUB). See Fig. \ref{fig:PUT} for an illustration. Here, the observed data are correlated with both the underlying utility features $\mathbf{U}$, which the encoder would like to convey to the receiver, and the sensitive features $\mathbf{S}$, which must be kept private. Quantifying both utility and privacy using mutual information, the trade-off between the two and the complexity of the shared information can be formulated as the following optimization problem:
\begin{equation}\label{CLUB_opt}
\mathop{\inf}_{\substack{ P_{\mathbf{Z} \mid \mathbf{X}}: \\ \left( \mathbf{U}, \mathbf{S}\right) \rightarrow \mathbf{X} \rightarrow \mathbf{Z}}} \mathrm{I}  \left( \mathbf{X}; \mathbf{Z} \right)  \; \quad  \; \mathrm{s.t.}  \quad  \mathrm{I} \left( \mathbf{U}; \mathbf{Z} \right) \geq R_{\mathrm{u}}, \;\; 
\mathrm{I} \left( \mathbf{S}; \mathbf{Z} \right) \leq R_{\mathrm{s}}. 
\end{equation}
where $\left( \mathbf{U}, \mathbf{S}\right) \rightarrow \mathbf{X} \rightarrow \mathbf{Z}$ indicates a Markov chain, while $R_{\mathrm{u}}$ and $R_{\mathrm{s}}$ are the lower bound on the utility of the shared information and the upper bound on the leaked private information, respectively. 

When the underlying data distribution is not known, ML techniques can be used to learn a privacy-preserving data release mechanism from data. Data-driven privacy mechanisms, inspired by GANs \cite{goodfellow2014generative}, model PUT as a game between a defender (privatizer) and an adversary \cite{huang2017context, osia2018deep, tripathy2019privacy}. The privatizer encodes datasets to minimize inference leakage on sensitive variables, while adversaries attempt to infer these variables from released data. 

In further research, PUT has also been extended to time-series datasets \cite{Shokri:ACM:10, Zhang:TIFS:19, Erdemir:TIFS:21}. For example, measurements of activity sensors shared with various health apps or social networking services, or location information shared with location-based services constantly, reveal sensitive information about a user's daily habits. In this case, each new data sample can provide additional information about the underlying sensitive feature, and simply obfuscating a single data sample or data over a fixed time interval does not suffice. Instead, all the released data history should be considered when deciding how to release new samples over time. In \cite{Erdemir:TIFS:21}, continuous data release problem is formulated as a Markov decision process, and various reinforcement learning techniques are deployed to solve it in practice. 

\subsubsection{Deep Joint Source-Channel Coding (DeepJSCC) for Semantic Communications}

When semantic communication takes place over a noisy communication channel, it can be formulated as a joint source-channel coding problem (JSCC) \cite{gunduz2024joint}. Although Shannon proved the optimality of designing source and channel coding schemes independently of each other in the information theoretic infinite blocklength regime, it is well known that such an independent design is suboptimal in the practical finite blocklength scenarios. Recently, pioneered by the work in \cite{bourtsoulatze2019deep}, data-driven JSCC schemes exploiting DNNs, known as DeepJSCC, have shown remarkable performance for general semantic communication tasks. Unlike traditional separate coding schemes, DeepJSCC integrates source compression and channel coding into a unified neural network architecture, enabling end-to-end optimization for semantic preservation. The main benefits of DeepJSCC can be listed as follows \cite{bourtsoulatze2019deep}: i) they can be trained on any target dataset and channel distribution, providing specialized designs for those scenarios; ii) they can be trained for any desired semantic communication task, as long as the corresponding loss function is differentiable (or can be approximately modeled as a differentiable function); iii) the resultant encoder/ decoder pairs have significantly less computational complexity compared to conventional source and channel coding schemes.  

Since DeepJSCC is an end-to-end coding paradigm, it can be combined with the PUT framework described in Section \ref{ss:PUT}, where the goal is to convey the input signal or some latent utility features of it to the legitimate decoder with the highest fidelity, while preventing the eavesdropper from inferring the private features. For example, continuing with the image transmission example above, one can aim at transmitting the image of a person to the legitimate receiver with the minimal distortion while making sure that the eavesdropper cannot infer certain private features of the person on the image, such as eye color, hairstyle, or gender. This problem was first studied in \cite{marchioro2020adversarial} and \cite{erdemir2022privacy}, where the authors employed GANs and variational autoencoders (VAEs), respectively. VAEs are particularly attractive in this context compared to standard autoencoders (AEs) used in most DeepJSCC solutions. VAEs embed the input signal to a distribution rather than a single point, and the corresponding channel input is sampled from a latent distribution rather than being generated by the encoder directly. In that sense, VAEs are better aligned with the stochastic encoding approach employed in information theoretic derivation of the wiretap channel capacity \cite{wyner1975wire}. The encoder in VAE-based DeepJSCC is modeled as a generative network, which allows not only the transmitter to control the latent distribution, but also the calculation of the variational approximations of the 
privacy leakage (based on mutual information) in a tractable manner. Another advantage of VAEs is that they can also sample discrete codewords from the latent distribution, which is difficult to achieve with conventional AEs due to non-differentiability. As previously reported in \cite{marchioro2020adversarial} and \cite{erdemir2022privacy}, these models can learn to align private features of the source signal with the channel subspace that cannot be decoded by the adversary, while other non-sensitive features are transmitted over the non-secure dimensions. In \cite{Letafati:TMLCN:25}, this work was extended to fading channels and multiple eavesdroppers.

Although the studies briefly reviewed above provide a certain level of security in semantic communications, they rely on various assumptions on the quality of the adversary's channel, or on the sensitive features that are being protected against leakage. On the other hand, JSCC solutions \cite{gunduz2024joint}, in general, have additional security challenges compared to their separation-based counterparts. The main benefit of JSCC compared to separation is that, in JSCC, the channel input is directly a function of the input signal. Therefore, no matter how noisy the underlying channel is, the channel output is always correlated with the input signal, and some estimate of the input signal can always be reconstructed. This results in the graceful degradation property of JSCC solutions--a property that is particularly attractive when communicating over high-mobility channels that are difficult to estimate. However, this property is also a vulnerability: channel output being correlated with the input signal always leaks information to potential eavesdroppers. Moreover, the analog nature of most JSCC schemes prevents the encoder from employing standard encryption techniques that require discrete alphabets.     

An alternative approach is proposed in \cite{tung2023deep}, which combines complexity-based encryption techniques with DeepJSCC solutions based on neural-networks. The scheme proposed in \cite{tung2023deep}, called deep joint source-channel and encryption coding (DeepJSCEC), is a public-key encryption scheme that leverages the learning with errors (LWE) problem introduced in \cite{Regev:ACM:09}. In particular, a specific definition of security, called ciphertext indistinguishability under chosen-plaintext attack (IND-CPA), is considered. In IND-CPA, rather than requiring that an eavesdropper learn absolutely nothing about the source signal, the goal is to guarantee that no probabilistic polynomial-time adversary can successfully recover the source information. Instead of mapping the input signal to continuous-amplitude channel input vectors through the encoder network, DeepJSCEC limits the transmitted channel symbols to a finite constellation as conventional digital communication systems, originally proposed in \cite{Tung:JSAC:22}. DeepJSCEC still relies on a pair of neural networks trained jointly for encoding and decoding, with the additional step of quantizing the channel codeword to points from the prescribed finite constellation before transmission over the channel. The finite constellation output is then encrypted before transmission. 

\begin{figure}[!t]
	\centering
	\includegraphics[width=3.4in]{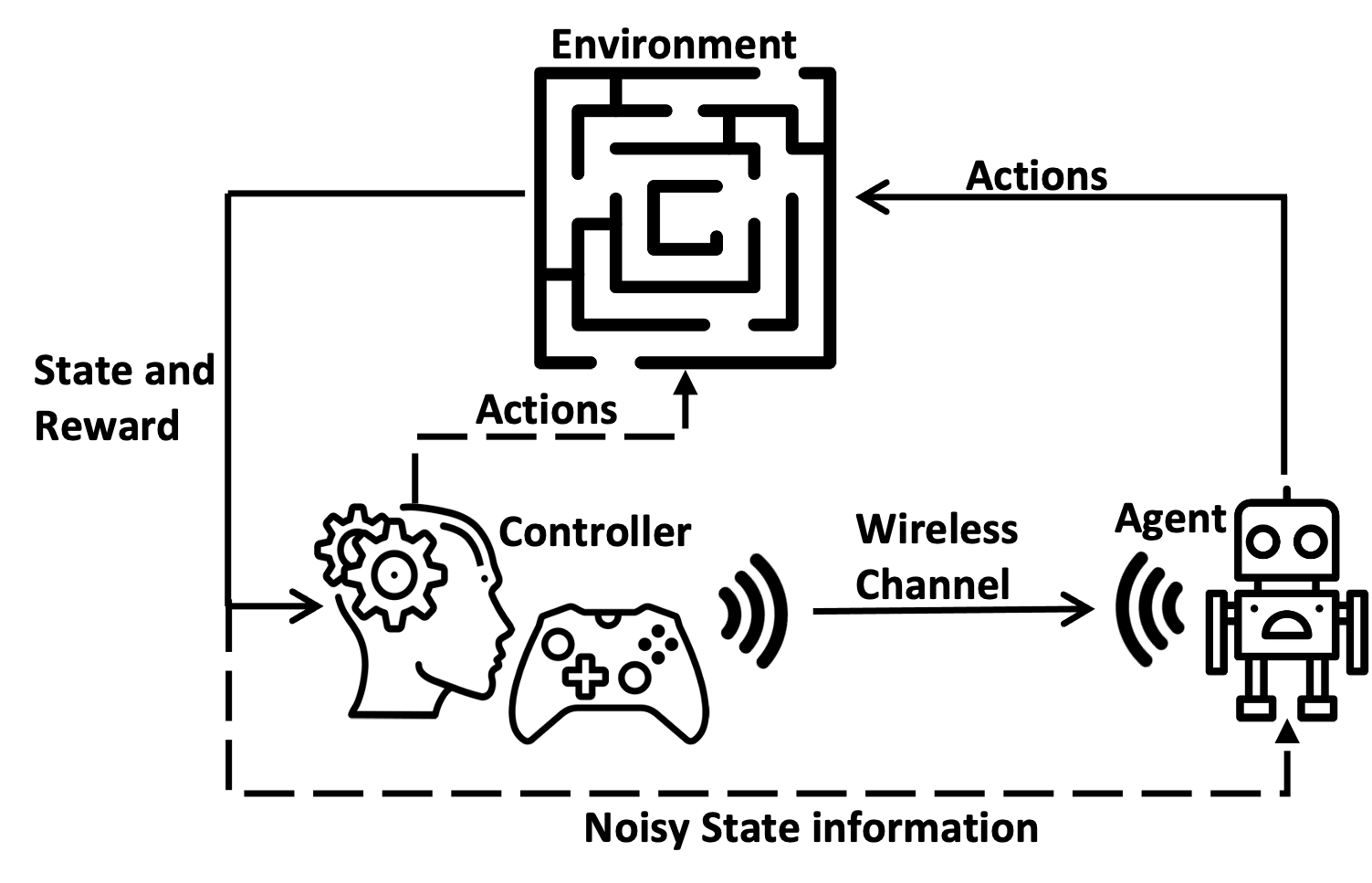}
	\caption{Illustration of a pragmatic/ goal-oriented communication system. Transmitter controls a remote agent through a noisy communication channel. Dashed lines may or may not be present. The goal is to maximize the average or discounted reward over a certain time horizon.}
	\label{pragmatic}
\end{figure}

\subsubsection{Security in Goal-Oriented/ Pragmatic Communications}

Semantic communication emphasizes the importance of considering the downstream task when communicating a signal over a channel. As explained in the previous section, this can still be considered in the general framework of rate-distortion theory or JSCC with a complex (e.g., non-additive) distortion function, which may also include distributional constraints as captured by the rate-distortion-perception trade-off \cite{hamdi2025perception}. However, in many cases, communication may have a goal that cannot be realized through a single round of transmission, and instead, the objective may be achieved through many interactions with the environment. In \cite{Gunduz:JSAC:23, Gunduz:BITS:23}, this has been formalized as pragmatic communications, where the receiver (or both the transmitter and the receiver) take actions in an environment, and the goal of communication is to maximize the average or discounted reward over many iterations. See Fig. \ref{pragmatic} for an illustration. 

Pragmatic communications can be considered in the context of multiple agents interacting in an environment with the objective of maximizing their overall reward. This is the setting of multi-agent Markov decision processes, where the agents may be sharing the same reward signal (i.e., cooperative team scenario), or have individual rewards (i.e., multi-player game scenario). Even in the case of a shared reward, when each agent has a partial/ noisy view of the environment, the problem becomes highly challenging as agents cannot coordinate their actions, and the environment becomes non-stationary. It these scenarios, additional communication among agents can help for coordination. It is known that communication allows agents to converge faster and to higher-reward policies \cite{Foerster:NIPS:16}. In this setting, agents need to learn not only how to interact with the environment, but also how to communicate among each other. This is known as \textit{emergent communications} \cite{Foerster:NIPS:16, jiang_learning_2018, jaques_social_2019, das_tarmac_2020}, where the agents learn a `language', that allows them to carry out a joint task. 

In Fig. \ref{pragmatic}, we illustrate a specific case with two agents as considered in \cite{Tung:JSAC:21}, where the controller has a perfect (or less noisy) view of the state, while the agent takes actions based on the signal received from the controller over a noisy channel. Differently from those studied in the ML literature, here the communication is noisy, and hence, the agents need to learn also to mitigate the effects of noise. This model is general enough to include many other problems studied in the literature as special cases; for example, those that involve age of information (AoI), or other related concepts \cite{roy_survey}. In these problems, the environment can be modelled as a Markov process observed by the controller whose state needs to be conveyed to a receiver over an error-free channel that may introduce random delays. In the general scenario introduced in \cite{Talli:JSAC:25}, the controller and the receiver agent may also take actions to optimize a generic reward function. 

Pragmatic communications introduces completely new sets of privacy and security challenges. As shown in \cite{Mason:arXiv:25}, even the timing of the messages transmitted over the channel can reveal information about the state of the environment to potential passive adversaries. In \cite{ufuk}, to defend against passive attacks, the authors limit the probability of an adversary observing the states to infer their transition probabilities under the chosen policy. A more active type of attack involves corrupting the observations of the controller to distort its knowledge about the system state and to derail its actions \cite{mo2009secure, smith2011decoupled, mo2010false, guo2017optimal}. The adversary can also corrupt the control signal \cite{bai2017data, zhang2016stealthy, kung2017performance, asaf2}. In the case of communication over a noisy or interference-prone channel, this may also be achieved by jamming or corrupting the channel. It is shown in \cite{Santi:ISIT:25} that by compromising the communication channel, an adversary can make the agent take actions that minimize its reward without being noticed by the controller.

%%%%%%%%%%%%%%%%%%%%%%%%%%%%%%%%%%%%%%%%%%%%%%%%%%%%%%%%%%%%%%%%%%%%%%%%%%
%%%%%%%%%%%%%%%%%%%%%%%%%%%%%%%%%%%%%%%%%%%%%%%%%%%%%%%%%%%%%%%%%%%%%%%%%%

\begin{figure}[!t]
	\centering
	\includegraphics[width=3.4in]{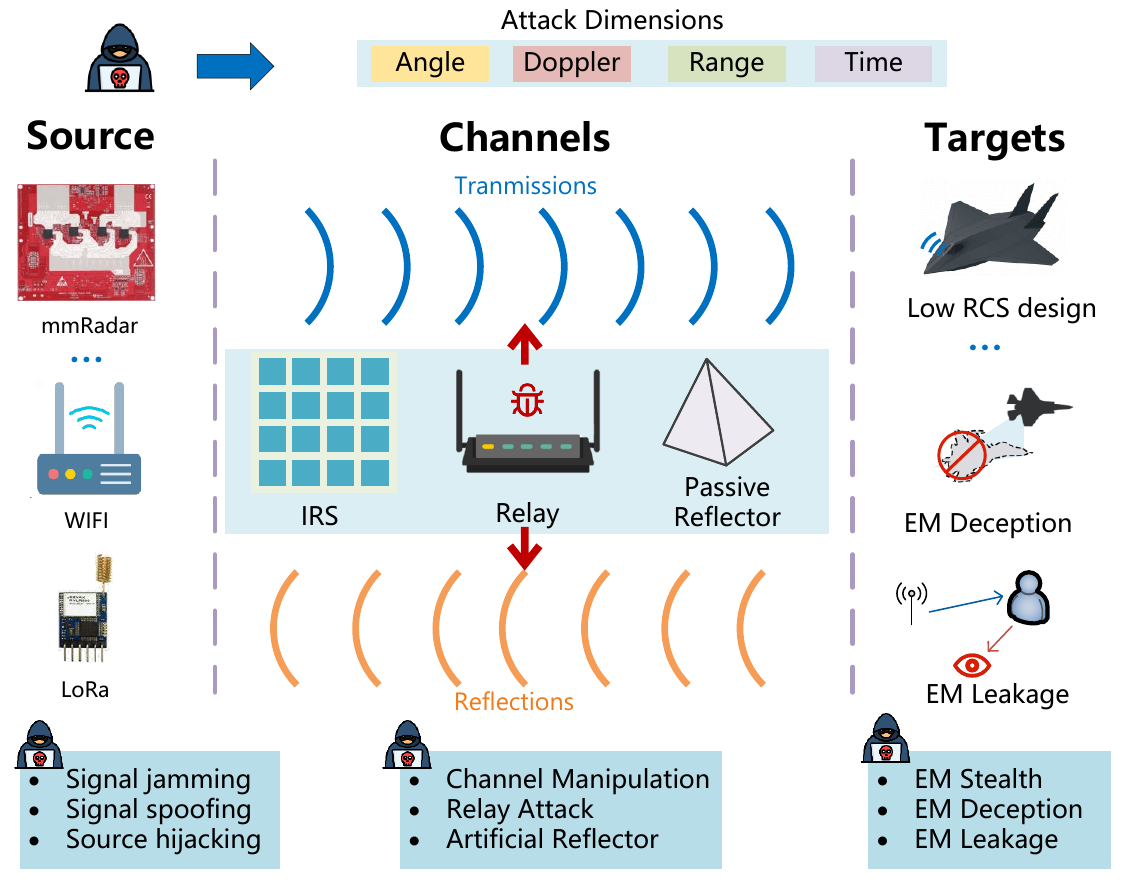}
	\caption{General framework for wireless sensing systems.}
	\label{Sensing}
\end{figure}

\section{Secure Sensing}\label{Sec:secure_sense}

%{\color{red}[Beixiong]}

Wireless sensing enables environmental perception by capturing variations in wireless signals as they interact with targets and the surrounding environment \cite{liu2019wireless,han2025rf,geng2024survey}.
Using propagation phenomena such as reflection, diffraction, and scattering, wireless sensing can extract implicit information about the state and behavior of targets. Such capabilities were previously used in a wide range of applications, ranging from healthcare monitoring and smart homes to industrial automation and human-computer interaction.
As illustrated in Fig. \ref{Sensing}, the transmitter emits radio frequency (RF) signals (e.g., WiFi, mmWave, LoRa, and RFID) that propagate through the physical environment and interact with surrounding targets.
By capturing and analyzing the echoed signals received at the receiver, the sensing system can detect the presence and/or estimate various physical characteristics, such as the target's position, orientation, speed, and material composition.
However, with the increasing prevalence of wireless sensing technology and due to the inherent broadcast characteristics of the wireless medium, security issues become more prominent.
Specifically, attackers can exploit vulnerabilities in wireless systems to compromise the accuracy of sensing systems through interference or deception of their key components.
In the following, we will elaborate on the main security issues in sensing systems from three perspectives: attacks on signal sources, wireless channels, and sensing targets, respectively. We then present forward-looking solutions addressing these challenges. 

\subsection{Security of Signal Sources}
\begin{figure}[!t]
	\centering
	\includegraphics[width=3.4in]{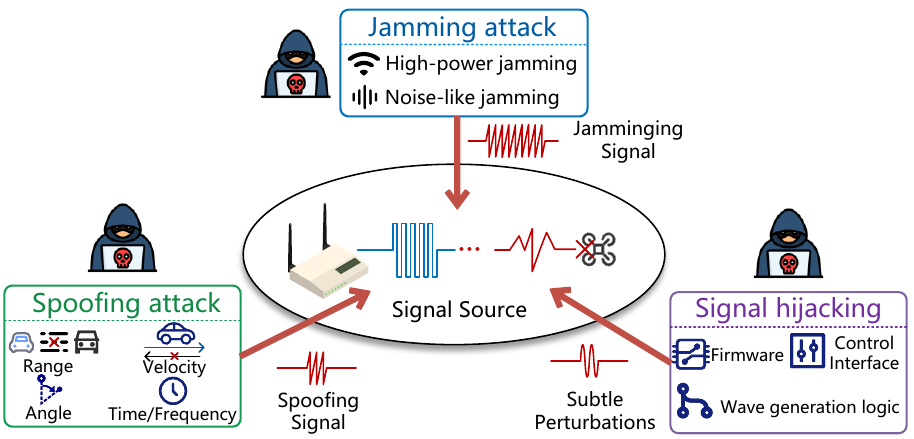}
	\caption{Attacks on signal sources.}
	\label{Source}
\end{figure}

Wireless sensing systems fundamentally rely on well-structured probing waveforms generated by the signal source to illuminate the surrounding environment and extract key physical information \cite{liu2019wireless}. 
Here, the signal source represents a critical vulnerability in the system.
Attacks on signal sources aim to disrupt the integrity and availability of the transmitted signals, degrading sensing accuracy, introducing false observations, or even disabling the entire sensing system.
Such attacks can be generally classified into the following three categories: jamming attacks, spoofing attacks, and signal source hijacking, each exploiting different aspects of the transmission process (see Fig. \ref{Source}). 

First, \textit{signal jamming attacks} degrade the system's sensing performance by injecting high-power or spectrally overlapping interference signals, reducing the SNR at the receiver. 
For instance, in mmWave automotive radar systems, single-frequency or swept-frequency jamming signals are carefully designed to exceed the SNR threshold of detection algorithms, thus effectively blinding the radar to target echo signals \cite{yan2016can}.
Besides the waveform-specific strategies, a more generalized form, \text{noise-like jamming}, involves emitting pseudo-random wideband signals that disrupt the sensing process regardless of the modulation scheme \cite{yan2016can,o2007introduction}. 
This technique is particularly effective in wideband and spread-spectrum systems, where it masks legitimate signals with spectrally similar interference, thereby complicating signal recovery and anomaly detection.
Although such jamming techniques can be highly disruptive, they demand accurate frequency alignment, directional synchronization, and sufficient transmission power, posing notable implementation challenges.

Second, \textit{signal spoofing attacks} aim to deceive the sensing system by transmitting counterfeit waveforms that closely replicate the legitimate signal in structure, timing, and spectral content \cite{geng2024survey}. 
These attacks exploit prior knowledge of system pilot structures or modulation patterns to inject well-aligned forged echoes. 
Depending on the estimated parameters, spoofing can be executed across different signal dimensions. For instance, range spoofing forges artificial time delays to mislead distance estimation \cite{Nallabolu2021AFrequency,hunt2023madradar,Miura2019ALR}; in its turn, velocity spoofing introduces Doppler-like frequency shifts to imitate target motion \cite{hunt2023madradar}; finally, angle spoofing manipulates signal phases to distort the perceived direction-of-arrival (DoA) \cite{Li2024SecBeam}.
%Depending on the estimated parameters, spoofing can be performed in multiple signal dimensions: range spoofing manipulates time delays, velocity spoofing introduces artificial Doppler shifts, and angle spoofing alters the received direction-of-arrival via phase manipulation. 
Notable real-time spoofing frameworks, such as mmSpoofing and MadRadar, demonstrate high-precision deception by dynamically estimating and replicating the sensing waveform \cite{vennam2023mmspoof,hunt2023madradar}.
Such attacks are not a simple signal generation behavior, but a highly complex attack method that relies on precise synchronization, accurate estimation, and environmental adaptability.

Third, \textit{source hijacking attacks} compromise the signal source internally by manipulating its firmware, control interface, or waveform generation logic \cite{geng2024survey}. 
Unlike external jamming or spoofing, hijacking represents an “insider threat” that directly undermines the trustworthiness of the waveform itself. 
Attackers may replace firmware, tamper with pseudo-random generators, or override system parameters, causing the signal source to emit malicious waveforms while appearing to operate normally \cite{geng2024survey}.
These attacks are particularly dangerous in software-defined radio (SDR)-based systems or cloud-controlled platforms where adversaries can exploit update channels or unsecured remote access points.

From a defensive standpoint, mitigating signal source-side attacks requires a multi-layered approach. 
Physical-layer protections include antenna shielding and signal filtering to suppress external interference \cite{han2025rf}. 
Furthermore, signal-level techniques, such as frequency hopping, time-varying modulation, or spread-spectrum coding, reduce susceptibility to jamming and spoofing by introducing unpredictability \cite{Wang2006Asurvey}. 
In addition, model-level defenses, including anomaly detection based on signal statistics or learning-based intrusion identification, can detect subtle deviations indicative of spoofing or hijacking \cite{ALSOUFI2024823}. 
Overall, trusted execution environments, firmware integrity checks, and secure boot processes are essential to defend against source hijacking and ensure waveform authenticity \cite{Liu2025ASurvey}.

\subsection{Security of Sensing Channels}
\begin{figure}[!t]
	\centering
	\includegraphics[width=3.4in]{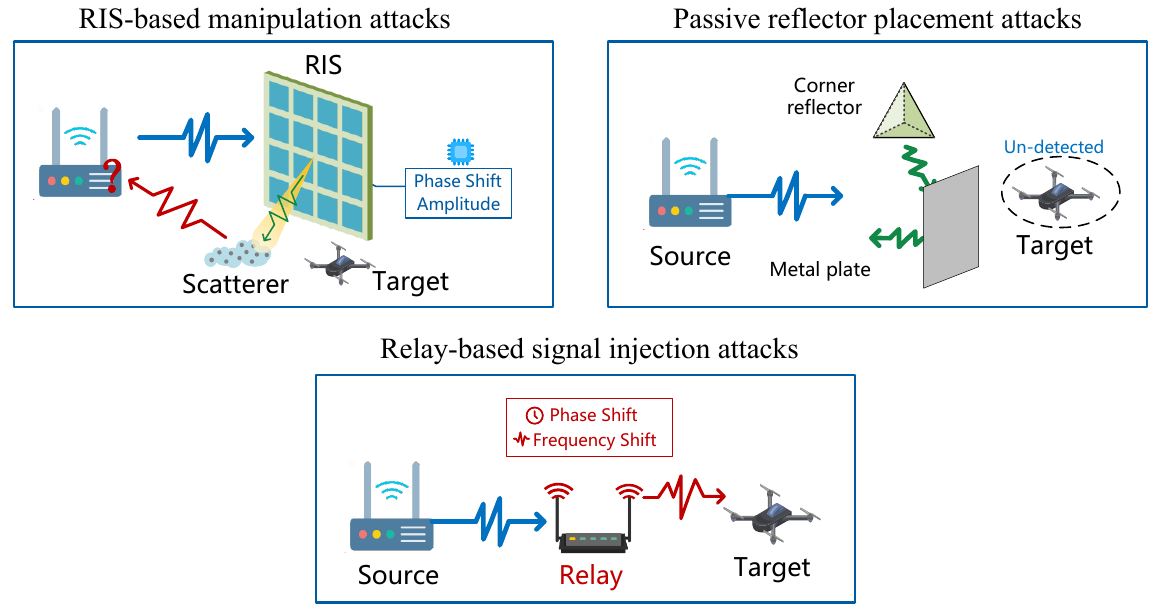}
	\caption{Attacks on sensing channels.}
	\label{channel}
\end{figure}
By analyzing channel-related signal features such as time-of-flight, angle of arrival (AoA),  and Doppler shift, wireless sensing systems extract crucial information about target's presence, location, and motion \cite{geng2024survey}.
However, the open and dynamic nature of wireless propagation makes the sensing channel an attractive attack surface.
Adversaries can manipulate propagation paths by forging, distorting, or concealing them, without directly interfering with the transmitter or the target being sensed \cite{Liu2025ASurvey}. 
As can be seen in Fig. \ref{channel}, this subsection introduces three representative types of channel-level attacks. To this end, RIS, relay devices, and passive reflectors, along with corresponding defense mechanisms, are used.

One prominent form of channel-level attack uses the \textit{IRSs} deployed in the surrounding environment \cite{zheng2025intelligent}.
When under adversarial control, IRS can be reconfigured to introduce artificial multipath components mimicking legitimate signal reflections.
By fine-tuning phase shifts, amplitudes, polarization of incident signals, attackers can generate false signal paths with specific delays or Doppler profiles, thus effectively creating phantom targets or simulating motion \cite{Liu2024AnIntelligent}.
Besides channel spoofing, IRS can also be used to suppress genuine reflections through destructive interference or introduce random channel variations that confuse tracking algorithms \cite{Yang2021ANovel}. 
These attacks are highly covert, as no active emissions are required, and their effects seamlessly blend into natural multipath phenomena, especially in cluttered or dynamic indoor environments.

By contrast, \textit{relay-based attacks} use active devices that receive, manipulate, and retransmit sensing signals \cite{Markantonakis2024Using}. These relays can introduce precise timing shifts to fake range estimates, apply frequency offsets to spoof velocity measurement, or employ beamforming techniques to mislead direction estimation. 
For instance, an amplify-and-forward relay may replay captured radar signals with a controlled delay, which makes the system to incorrectly determine the target's position \cite{Staat2022Analog}. 
While relay-based attacks offer greater flexibility in modifying signal properties, their active transmission behavior also increases the likelihood/risk of detection as compared to passive IRS manipulation.
A more basic, yet effective method involves the placement of \textit{passive reflectors}. Adversaries can deploy static objects, such as metallic sheets, engineered scatterers, or corner reflectors \cite{Zhu2023TileMask}, at strategic locations to generate misleading multipath reflections.
These additional reflections may appear indistinguishable from real targets, confusing localization systems or artificially increasing target numbers.
In structured environments with predictable signal propagation, such as warehouses or tunnels, passive reflectors are particularly deceptive due to their simplicity, low cost, and the absence of electronic signatures.

To mitigate these threats, advanced defense strategies must be implemented. For IRS-based attacks, techniques such as angular diversity, randomized IRS probing, or channel consistency checks can detect abnormal reflections deviating from expected propagation pattern \cite{Asaad2022Secure}.
Relay-based attacks can be countered using secure timing protocols or waveform fingerprinting, which validate physical plausibility of received signals \cite{Vu2022AComparative}. 
For passive reflector scenarios, environment-aware sensing, use of prior knowledge of the map, sensor mobility, or spatial correlation analysis can identify static anomalies that are inconsistent with expected target behavior \cite{Chorti2022Context}. 
These cross-layer defensive mechanisms enhance system resilience and ensure reliable sensing performance in adversarial environments.

\subsection{Security of Sensing Targets}
\begin{figure}[!t]
	\centering
	\includegraphics[width=3.4in]{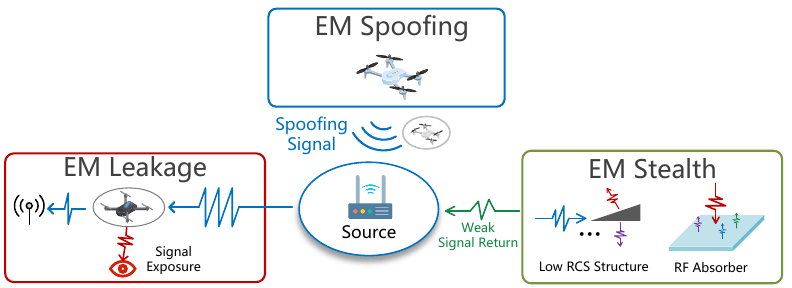}
	\caption{Attacks on sensing targets.}
	\label{Target}
\end{figure}

%\begin{figure}[it]
%	\centering
%	\includegraphics[width=3.4in]{Target.pdf}
%	\caption{Attacks on wireless signal targets.}
%	\label{Target}
%\end{figure}
From the perspective of sensing targets, adversarial threats emerge when the physical or electromagnetic properties of a target are intentionally manipulated to evade detection, mislead sensing results, or leak private information \cite{geng2024survey}. 
These threats can be broadly categorized into the following three types: 1) electromagnetic (EM) stealth; 2) electromagnetic spoofing; and 3) electromagnetic leakage (see Fig. \ref{Target}). Each type reflects a different adversarial objective and mechanism and poses distinct challenges to the integrity of sensing systems.

First, \textit{stealth attacks} aim to conceal the presence of the target by suppressing its electromagnetic signature.
Traditional methods include geometry shaping, where target contours are designed to deflect incident EM waves away from sensing devices, thus reducing radar cross-section (RCS) and detectability \cite{shin2021shape}. 
Another common approach involves EM wave-absorbing materials or stealth coatings that attenuate backscattered energy in specific frequency bands \cite{wang2017radar}. 
However, both aforementioned approaches face limitations in adaptability and performance under dynamic or multi-view sensing environments.
To overcome these constraints, several recent studies explored the use of IRS-mounted on targets, which can dynamically adjust reflection phase and direction to cancel or redirect radar echo signals \cite{zheng2025intelligent,Zheng2024Intelligent,Xiong2024ANew,Wu2025Intelligent}. 
Compared to passive materials, IRS-assisted stealth offers programmable, real-time control, and improved concealment under complex scenarios.

Second, \textit{spoofing attacks} seek to mislead the sensing system by generating false electromagnetic responses mimicking those of real targets \cite{geng2024survey}. On the physical level, decoy-based spoofing employs towed reflectors, chaff, or corner reflectors to introduce misleading echoes and clutter around actual targets \cite{Zhou2011Performance}.
More advanced implementations use modulated metasurfaces that embed frequency- or time-varying responses into passive structures, thus enabling the generation of multiple controllable echoes with distinct delay, Doppler, or angle characteristics \cite{Fang2024Diverse}. 
These false signals can be designed to imitate legitimate targets while remaining passive and difficult to trace.
To further enhance deception in dynamic environments, IRS-based spoofing is used to dynamically alter reflected signals to resemble those from synthetic or displaced targets, without emitting any new waveforms \cite{zheng2025intelligent,Wang2024Intelligent}. 
Such attacks significantly complicate target localization and classification for authorized sensing devices.

Third, \textit{electromagnetic leakage} refers to the unintended exposure of sensitive information through the target's natural interaction with wireless signals \cite{Ramesh2022TickTock,Zou2023IoTBeholder}. 
Passive adversaries can analyze subtle changes in CSI or received signal strength indicators (RSSI) to infer presence, movement, gestures, or even physiological states such as respiration or heartbeat \cite{Li2024Practical}.
This type of attack poses significant privacy concerns, particularly in smart homes, healthcare monitoring, and human-computer interaction scenarios. 
Since the target itself does not actively emit signals, these attacks are difficult to detect and frequently operate without violating traditional security boundaries.

Protecting sensing targets requires a multi-faceted approach. To defend against stealth attacks, sensor fusion, multi-angle illumination, and adversarial-aware detection algorithms can help to identify suspicious signal suppression \cite{MIAO2025104149}. 
Spoofing defenses rely on cross-domain verification, consistency checks between time, frequency, and angle signatures, and anomaly detection in motion patterns \cite{Liu2025ASurvey}. 
To mitigate electromagnetic leakage, privacy-preserving sensing protocols, spatial obfuscation techniques, and signal randomization can be applied to reduce the risk of semantic inference from passive eavesdropping \cite{Nguyen2021Security}. 
With the ongoing evolution of sensing capabilities, it is crucial to develop target-centric security designs that would balance functionality, usability, and protection.

%%%%%%%%%%%%%%%%%%%%%%%%%%%%%%%%%%%%%%%%%%%%%%%%%%%%%
\section{Secure Coded Computing}\label{Sec:secure_compute}
Secure computation is an emerging pillar of 6G networks, alongside secure communications and sensing. The vision is that future intelligent wireless systems will not only transmit and sense data securely, but also compute on distributed data in a secure and privacy-preserving manner. This is driven by the rise of edge/cloud computing and federated learning (FL) in 6G, where large-scale data processing is offloaded to untrusted nodes or collaborative devices. Traditional cryptographic approaches (e.g., fully homomorphic encryption \cite{FHE}) enable computing on encrypted data but remain impractical for real-time wireless applications due to high complexity. Coding- and information-theoretic methods offer novel ways to achieve secure, robust, and efficient distributed computations \cite{OurSecureFunctionCompt,ITSecTutorial}. For instance, adding structured redundancy can make computations resilient to errors or stragglers (i.e., slow/unresponsive nodes), and encoded computations can hide sensitive data from curious computing nodes. Many recent theoretical advances in coded secure and private computing have not yet been translated into practice, creating a pressing need for research that bridges this gap. At the cusp of an AI-driven era, ensuring efficient and secure computations on large datasets using unreliable or untrusted components are critical. Thus, 6G must enable distributed intelligence (i.e., training, inference, and data analytics) with guarantees of data privacy, integrity of results, and robustness against system uncertainties. In practice, these systems must operate under heterogeneous threat models, including honest-but-curious, malicious, and colluding computing nodes. 

This section surveys selected key approaches for secure coded computation, focusing on techniques beyond secure communication or sensing, and highlights future directions; see, for instance, \cite{CDCSurvey1,Securecomputingsurvey2,Securecomputingsurvey3} for extended discussions about secure computing.

%%%%%%
\subsection{Coded Distributed Computing for Robust and Secure Computation}
A cornerstone technique for secure distributed computation is coded computing, which injects redundancy into computational tasks to achieve both straggler tolerance and secrecy. In classical coded distributed computing (CDC) frameworks \cite{CDCfirst}, function computation tasks are split into subtasks and encoded, such that a central node can recover the final result from any sufficiently large subset of worker outputs. This coded approach reduces the communication load of intermediate data shuffling by a factor related to the computation redundancy and ensures that the computations are carried out despite some slow or failing workers. Straggler mitigation is vital in wireless edge computing, where node computation speeds vary. For example, coding a matrix multiplication into $n$ pieces, so the result can be decoded from any $k<n$ completed pieces can eliminate waiting for the slowest nodes \cite{StragglerCDC}. Moreover, security can be layered into these codes by using techniques like polynomial secret sharing or noise injection on encoded subtasks. In this way, each worker receives an encoded fragment that, by itself, reveals no sensitive information even if some workers or eavesdroppers collude. Advanced techniques that rely on the properties of discrete Fourier transform \cite{DenizSecureDMM} or bivariate polynomial coding \cite{Hasircioglu:JSAC:22} provide state of the art results in terms of the communication load necessary to guarantee perfect security against workers and the required level of straggler robustness.

Recent research in \cite{OursecureDC} shows that secure CDC schemes use novel codes and customized secret sharing that protect data against powerful adversaries (including quantum-capable eavesdroppers, which is possible due to information-theoretic security guarantees). Such schemes, applied to tasks like distributed matrix multiplication, guarantee that nothing about the original matrices leaks from the information exchanged, while incurring minimal extra overhead. Moreover, coded computing frameworks have been generalized to complex network topologies, such as multi-layer multi-access computing models with separate mapper and reducer nodes \cite{OurMultipleAccessCDC}, without losing the straggler mitigation and security benefits. In summary, secure CDC provides a resilient and secure processing layer for 6G and combats random delays, failures, and eavesdroppers. This illustrates how redundancy can simultaneously support reliability and information-theoretic security.

%%%%%%%
\subsection{Privacy-Preserving Distributed Learning and Aggregation}
A prominent application of secure computation in 6G is FL \cite{FederatedLearning}, where many wireless devices collaboratively train a model without sharing raw data. The main challenge is to aggregate local model updates at a central server privately and accurately. One solution is secure aggregation, which allows an aggregator to compute the sum of user updates without seeing any individual update \cite{SecureAggregation}. Information-theoretic techniques underlie such protocols and ensure that an honest-but-curious server (or any eavesdropper) only obtains the aggregate result. Secure aggregation has become the de facto approach in industry deployments of FL (e.g., Google’s Gboard \cite{GboardFL}). Another approach to client privacy is local differential privacy (LDP), where each device adds calibrated noise to its data or model update before transmission \cite{DworkDP}. Recent work has studied the combination of LDP with communication-efficient training, showing how quantization and noise impact learning performance \cite{LDPwithchannelsynthesis}. Such privacy-preserving distributed learning methods must also cope with Byzantine adversaries (i.e., malicious participants sending counterfeit updates to corrupt the model). Robust distributed learning can leverage ideas from coded computing and robust statistics to counteract malicious updates. For instance, Byzantine-resilient gradient coding adds redundant computations so that the server can detect or tolerate a fraction of counterfeit updates \cite{Byzantinesecuregradientcoding}. The overarching goal is to enable privacy-preserving and robust distributed learning in 6G such that (i) users' personal data, including health or behavioral data, remain local, (ii) model updates are aggregated securely, and (iii) the training process is resilient to unreliable or malicious devices. This is an active area of research and standardization, with new codes and protocols being developed for private FL.

Another approach to private distributed learning in wireless networks is to exploit the inherent noise and interference in the wireless channel for privacy. We note that, at its core, computing over a wireless network is a distributed JSCC problem \cite{gunduz2024joint}, where the goal is to compute the result of a function with minimal distortion over a noisy multiple access channel. It is well known that the optimality of separation breaks down in this setting even in the usual information theoretic asymptotic formulation. The superposition property of the wireless medium is exploited in \cite{Amiri:TSP:20, Amiri:TWC:20} to achieve efficient over-the-air computation (OAC) by sending model updates in an uncoded fashion. In this scheme, the sum of model updates is received at the server with additional channel noise, which can provide a certain level of differential privacy (DP) for the final trained model. In \cite{seif2020}, if the channel noise is not sufficient to satisfy the desired DP target, some of the devices transmit additional noise, benefiting all devices. Instead, in \cite{Koda2020DifferentiallyPA} and \cite{Liu:JSAC:21}, the transmit power is adjusted for the same privacy guarantee, but this relies on the perfect knowledge of the channel gain. %The authors in \cite{seif2021privacy} showed that jointly optimizing both wireless aggregation and user sampling can further improve differential privacy.
Instead, in \cite{hasircioglu2021private}, the authors exploit the anonymity provided by OAC for privacy, which prevents the server from detecting which devices are participating in each round.

%%%%%%%%%
\subsection{Function Computation Privacy and Advanced Secure Coded Computing Paradigms}
Beyond protecting data privacy, future networks may also require protecting the privacy of the function or task being computed. In some scenarios, even knowing which function is being evaluated on the data (or how tasks are assigned to nodes) can leak sensitive information or business intelligence. Private coded computing addresses this by concealing task assignments and computation targets. A recent model in \cite{OurPrivateDC} introduces privacy constraints into CDC such that the mapping of subtasks to workers is hidden. Using an extended placement delivery array (PDA) construction, originally used in coded caching problems \cite{PDAinCaching}, they characterize the fundamental trade-off, where enforcing function privacy inevitably increases the communication load but optimized coding strategies can significantly alleviate the overhead. Here, workers perform computations without knowing which functions are computed by other workers. This line of work yields information-theoretic privacy guarantees. Another frontier is to leverage semantic communication concepts for secure computation. Randomized distributed function computation (RDFC) framework, introduced in \cite{MyRDFC,DidrikISIT2025}, computes randomized function outputs by using coordinated random codebooks that exploit the semantic structure of the data. By allowing the decoder output to be a randomized function of the inputs, one can drastically reduce the communication load compared to classical compression methods, while guaranteeing strong security/privacy for each input data point. This approach combines semantic communication and strong coordination \cite{CuffChannelSynthesis} principles to simulate correlated randomness between a transmitter and receiver, so the receiver obtains a function output that is statistically consistent with a desired secure/private computation. Early results show that semantic compression techniques can outperform standard lossless approaches in secure and private function computation, indicating a promising direction for 6G where communication and computation resources are jointly optimized. Although many of these advanced techniques are still in nascent stages, they form a toolbox of emerging paradigms, from function-private computing to semantic security, that will shape secure computation in 6G. 

These developments indicate that coding-theoretic methods will remain central to scalable and trustworthy distributed computing in 6G.

%%%%%%%%%%%%%%%%%%%%%%%%%%%%%%%%%%%%%%%%%%%%%%%%%%%%%%  GK part 
\section{Unified Security Design for Future Networks}\label{Sec:unified_design}

\subsection{Adversarial Attacks in Wireless Networks: Challenges and Design Insights}

The increasing reliance on ML in wireless communication systems introduces novel vulnerabilities, notably adversarial attacks. These attacks craft small, frequently imperceptible perturbations to the input signals in order to mislead learning-based models. In the context of wireless networks, such attacks can drastically degrade the performance of signal classifiers, modulation recognizers, and channel estimation models, thereby threatening reliability and security of the system.

\subsubsection{Adversarial Attack Formulation and Threat Model in Wireless Networks}
Depending on the adversary's objectives and constraints, adversarial attacks in machine learning can be approached through different optimization strategies. Two primary formulations widely used in the literature include the best-effort and minimum-norm formulations. Both aim to craft perturbations that alter the model's predictions; however, they fundamentally differ in how they balance effectiveness and resource usage.

The best-effort formulation is commonly adopted when the attacker operates under a fixed perturbation budget. In this scenario, the goal is to maximize the impact of the adversarial perturbation without exceeding a predefined limit on its strength. This approach is particularly useful in practical settings where maintaining imperceptibility or adhering to power constraints is crucial—for instance, in stealthy evasion attacks that must bypass detection systems. Best-effort attacks are also favored in comparative studies, as the fixed constraint provides a consistent metric for the evaluation of the relative performance of different attack methods.

By contrast, the minimum-norm formulation is designed for scenarios where the success of the attack is prioritized over perturbation size. Here, the objective is to find the smallest possible perturbation guaranteed to fool the model. This formulation is particularly relevant in robust regions of the model's decision space, where small perturbations may be ineffective due to high confidence in the original prediction. This formulation is well-suited for evaluating the model’s resilience, identifying the minimal conditions under which its behavior changes. Despite being more computationally demanding, this approach ensures that the crafted adversarial sample will achieve the desired effect, making it ideal for security-critical applications or in the analysis of model vulnerabilities.

In summary, best-effort attacks focus on achieving the most with limited resources, which makes them practical and efficient for benchmarking and stealthy exploitation. On the other hand, minimum-norm attacks emphasize reliability and success, even at the expense of increased perturbation size, thus offering a deeper understanding of model robustness under worst-case conditions. Complementing each other, both formulations are used according to the specific goals and constraints of the adversarial scenario.

\subsubsection{Taxonomy of Adversarial Attack Methods}
Adversarial attacks in machine learning are commonly characterized along several dimensions, all of which can be formally captured within an optimization-based adversarial problem. For instance, the adversary's knowledge of the target model determines whether the attack is white-box (i.e., full access to model internals), gray-box (i.e., limited access to outputs), or black-box (i.e. minimal prior knowledge, frequently treated as theoretical due to practical constraints). These distinctions influence the attack formulation—while white-box attacks use gradients directly, black-box attacks resort to query-based approximations. In addition, adversarial attacks differ in their purpose: while evasion attacks manipulate inputs during inference to cause misclassification, poisoning attacks corrupt training data to compromise model behavior, Trojan attacks plant hidden triggers for future misuse, while inference-based attacks aim to extract sensitive model information or training data membership. Each of these goals determines a different attack strategy, frequently unified under constrained optimization frameworks embedding both the attacker's capabilities and objectives.

Further distinctions arise from the time of impact (training-time, inference-time, or both), specificity of the outcome (targeted vs. untargeted misclassification), and perturbation mechanism (gradient-based vs. non-gradient-based). Depending on access levels, attacks can also operate in different spaces: either in the physical domain (e.g., tampering raw inputs like images or signals) or in the feature space (e.g., altering internal representations). In wireless communication contexts, the attack location introduces further complexity specifically, while local attacks affect the transmitter or receiver (e.g., via malware), over-the-air attacks require real-time interference synchronized with legitimate transmissions. Finally, adversarial goals are aligned with the classical confidentiality, integrity, availability (CIA) triad: compromising confidentiality (e.g., through Trojans or spoofing), integrity (via targeted evasion), or availability (by degrading model performance through poisoning or demodulation disruption). All these dimensions, traditionally expressed as a "threat model," can in fact be fully described by specifying the optimization problem governing the attack—making the adversarial problem a more general and formal definition of the threat landscape.

\subsubsection{General View on Adversarial Attack Methods}

The field of adversarial machine learning was initiated with the introduction of the limited-memory Broyden–Fletcher–Goldfarb–Shanno (L-BFGS) attack. This attack demonstrated that small, carefully crafted input perturbations could significantly alter a deep neural network’s prediction. Although initially proposed to explore model interpretability, this approach revealed that the loss landscape of models trained via empirical risk minimization (ERM) contains regions vulnerable to optimization-based exploitation. As a result, L-BFGS became the first successful white-box evasion attack, showcasing that optimization techniques could directly craft adversarial examples.

Building on this foundation, the fast gradient sign method (FGSM) introduced a more computationally efficient alternative. Instead of solving complex optimization problems, FGSM uses gradient information to construct a single-step adversarial perturbation maximizing the model’s loss. Its simplicity and speed made it a cornerstone method for evaluating white-box robustness. Later improvements such as the basic iterative method (BIM) and momentum iterative method (MIM) enhanced FGSM by applying perturbations over multiple steps. These iterative methods enabled a finer control and stronger attacks, particularly in scenarios where one-step attacks such as FGSM were insufficient. The projected gradient descent (PGD) attack further extended this approach by projecting perturbations onto a constrained space, making it one of the most widely used and benchmark methods in adversarial robustness research.

While gradient-based attacks tend to affect many features, a different paradigm was introduced by one-pixel attacks such as the Jacobian-based saliency map attack (JSMA). Instead of spreading perturbation across the input, JSMA identifies and modifies only the most influential input dimensions—frequently,  just one or a few pixels. This selective manipulation targets specific neurons responsible for class confidence and provides a fundamentally different threat model. Importantly, defenses effective against FGSM-like methods may be ineffective against JSMA, which highlights the need to consider both classes of attacks when evaluating robustness.

Furthermore, the Carlini \& Wagner (CW) attack marked another milestone by proposing a flexible framework formulating the adversarial objective using surrogate functions, thus enabling fine-grained control over perturbation norms and model outputs. Its effectiveness across various settings made it a standard benchmark for the evaluation of defense mechanisms. In parallel, the universal adversarial perturbation (UAP) strategy explored perturbations that generalize across many inputs. Instead of crafting one sample at a time, UAP seeks a single perturbation that can fool a broad set of inputs, thus demonstrating strong transferability—even across different models—which makes it particularly relevant in black-box settings.

More recent developments focused on adaptive and distributional strategies. For instance, auto-PGD (APGD) attack enhances PGD by incorporating dynamic step-size adjustment and momentum, thus preventing the attack from getting stuck and improving convergence towards optimal adversarial directions. Due to its reliability and efficiency, APGD is frequently used as a standard method for robust evaluation. Finally, by generating adversarial distributions rather than individual samples, the distributionally adversarial attack (DAA) introduces a novel conceptual shift. Via optimizing over distributions of inputs rather than point-wise examples, DAA captures a more global view of model vulnerability and better simulates worst-case generalization risk.

Furthermore, in \cite{PNGM}, the authors proposed the projected natural gradient descent (PNGD) attack. It is an adversarial attack method that perturbs inputs by following the natural gradient direction, which accounts for the geometry of the model's parameter space. In contrast to standard gradient methods, PNGD projects the perturbation back into a constraint set to ensure the adversarial example stays within a bounded region. This approach leads to more effective and efficient attacks, particularly against models robust to conventional gradient-based perturbations.

In summary, from simple, interpretable  perturbations, adversarial attacks have evolved to complex and adaptive strategies. Each method serves distinct goals—whether optimizing speed (FGSM), effectiveness (CW, APGD), stealth (JSMA), or generalization (UAP, DAA)—and reflects the growing sophistication in modeling both the attacker’s capabilities and the model’s weaknesses. Understanding these differences is essential for the development of comprehensive defenses and robust learning systems.

\subsubsection{Adversarial Attacks on Wireless Communication}

\paragraph{Adversarial attacks on communication}
In \cite{AMCad12}, the authors employed the fast gradient method (FGM) to craft adversarial perturbations for the vectorized time-domain CNN 2 (VT-CNN2) modulation classifier. The authors identified vulnerable classes through principal component analysis (PCA) and demonstrated that minimal perturbation power is sufficient to significantly reduce classifier accuracy. Similarly, in \cite{AMCad15}, the authors compared FGSM and L-BFGS attack methods, showing that L-BFGS requires less perturbation power and is more effective in adversarial scenarios, particularly under normalization constraints. The aforementioned analysis also included the impact of adversarial perturbations in the frequency domain, thus reinforcing the need to evaluate signal-domain effects in automatic modulation classification (AMC) robustness.

However, in \cite{AMCad12, AMCad15}, the authors assumed that they were able to compromise directly the eavesdropper’s received signal ignoring the effect of the channel on the perturbed signal, which limits the applicability of these attacks in practice. In \cite{Hameed:TIFS:21}, the authors considered modifications directly on the transmitted symbols. Moreover, they also assumed that the intended receiver is oblivious to the modifications employed by the transmitter to confuse the intruder; and therefore, the goal of the transmitter is to introduce modifications to the transmitted signal sufficient to fool the intruder, but not exceeding than the error correction capabilities of the intended receiver. The results revealed that, through curriculum learning across different channel SNRs, intruder can become more robust against both the channel noise and the defensive
perturbations. The authors also found that reliable communication can still be achieved by reducing the communication rate while preventing the intruder from detecting the modulation used. 

Furthermore, in \cite{RMad8}, the authors developed a grey-box adversarial framework alternating between evasion and poisoning attacks based on traffic prediction outcomes derived from a surrogate model. This method enables adaptive attacks on spectrum sensing phases. Complementarily, in \cite{RMad5}, the authors presented a poisoning strategy that maximizes loss over clean data by retraining on adversarially perturbed examples. The attack is cast as a bi-level optimization problem and demonstrates severe degradation in model performance, particularly in FL setups. To mitigate such threats, defensive mechanisms were proposed.

Furthermore, in \cite{secad1}, the authors proposed a GAN-based spoofing attack framework against wireless intrusion detection systems (IDS), where the adversarial generator crafts synthetic inputs that deceive the IDS discriminator. This method was found to achieve a high spoofing success even under location variance. In a related effort, \cite{secad5} introduced intrusion detection system generative adversarial network (IDSGAN), a grey-box adversarial learning framework generating malicious traffic records to evade IDS detection. The generator was trained using feedback from the IDS, allowing it to adapt without requiring full access to the IDS model, which makes it highly effective in practical grey-box scenarios.

In \cite{MIMOad2}, the authors investigated targeted white-box adversarial attacks against deep-learning-based MIMO communication systems. Here, by crafting input perturbations with full access to the model parameters, the attacker misleads the receiver's classifier to predict specific incorrect classes. The proposed attacks use gradient information and demonstrated significant performance degradation under minimal perturbation power. The results underscored the need for robustness analysis in MIMO receivers that integrate deep neural networks.

In \cite{risad}, the authors designed targeted adversarial attacks for RIS-assisted wireless signal classification systems. Using controllable reflection properties of RIS, gradient-based perturbations that modify the received signals so as to misguide the classifier were applied. The results revealed that, while enhancing communication performance, RIS can inadvertently expand the attack surface when integrated with ML models, thus necessitating RIS-aware robustness mechanisms.

In another relevant study \cite{UAVad1}, the authors presented a targeted adversarial attack on unmanned aerial vehicle (UAV)-based wireless systems where the attacker crafts gradient-based perturbations that force misclassification of received signals. The attack maintains the signal's spectral structure, making it stealthy and practical in airborne applications. In \cite{NTCad2}, an untargeted attack scenario was analyzed, showing that non-terrestrial communication (NTC) systems, such as satellite or UAV-based classifiers, are vulnerable to input perturbations that degrade overall classification accuracy. These attacks underscore the need for robust AI integration in critical non-terrestrial platforms.

Furthermore, adversarial machine learning has recently been applied to semantic communication systems, which rely on end-to-end deep learning architectures to transmit and reconstruct meaning, rather than raw symbols. In several previous studies \cite{E2Ead}, \cite{E2Ead2}, and \cite{E2Ead3}, the authors proposed white-box evasion attacks targeting autoencoder-based semantic communication pipelines. The results of these studies revealed that minor perturbations in the input or latent semantic space can considerably impair semantic reconstruction accuracy at the receiver, even under benign channel conditions. Notably, \cite{E2Ead3} demonstrated that channel-aware perturbations crafted at the transmitter can propagate through the entire semantic pipeline, resulting in a critical information loss. These attacks highlight the need for robust semantic-level encoding strategies and adversarial training techniques to protect against subtle, yet harmful perturbations in future semantic-aware wireless systems.

\paragraph{Adversarial attacks on integrated communication and sensing}
ISAC is a core component of 6G aimed at unifying radar sensing and wireless communication within a single framework. However, along with providing greater spectrum efficiency and reduced hardware costs, this integration also introduces new vulnerabilities to adversarial attacks. Many ISAC systems employ deep neural networks for radar-based perception tasks such as object detection and localization. For instance, in \cite{liu2023adversarial}, the authors demonstrated that imperceptible perturbations can significantly degrade the performance of such radar classifiers. Similarly, in \cite{hu2022security}, the authors highlighted the risks of adversarial inputs in vehicular ISAC systems, where small perturbations may lead to incorrect scene understanding and compromised safety. Beyond digital attacks, physical-layer manipulation also poses a severe threat. In \cite{chen2023physical}, the authors showed that carefully placed physical artifacts can create ghost targets in radar scenes, misleading ISAC perception without requiring digital access. In \cite{gong2022security}, the authors discussed RF fingerprint spoofing attacks that disrupt indoor localization systems based on ISAC principles. Furthermore, the authors of \cite{zhang2023ai} analyzed cross-layer adversarial attacks in AI-native 6G architectures, where corrupted sensing outputs lead to faulty decisions in beamforming and scheduling. Taken together, these findings underline the critical need for robust AI models, resilient sensing algorithms, and multi-layer defense strategies to ensure secure ISAC deployment in 6G networks.

\paragraph{Adversarial attacks on computing in 6G wireless networks}
The transition towards AI-native 6G architectures introduces pervasive computing across the network, particularly through edge intelligence, distributed learning, and on-device inference. However, while these advancements enhance performance and scalability, they concurrently expose the wireless infrastructure to adversarial threats targeting the computing components of the network. Furthermore, edge devices and base stations running deep learning models are susceptible to adversarial examples—carefully crafted inputs that mislead model predictions. In \cite{bair2022adversarial}, the authors surveyed adversarial machine-learning threats in wireless systems and emphasized the vulnerability of RF deep learning classifiers and edge-based modulation recognition models. In another relevant study \cite{shlezinger2021model}, the authors highlighted the integration of model-based and data-driven computing in 6G, warning that adversarial noise in learned components could lead to incorrect signal reconstruction and resource allocation. Similarly, in \cite{zhang2023crosslayer}, the authors explored cross-layer adversarial attacks, where malicious inputs on the physical layer impact higher-layer functions such as traffic scheduling and power control, propagating failures throughout the AI-based computing stack. Moreover, in \cite{li2021federated}, the authors discussed vulnerabilities of FL in wireless environments, where adversarial clients can poison model updates or inject backdoors into collaboratively trained models. Together, these findings point to the urgent need for robust FL, certified defenses, and secure model deployment mechanisms in 6G systems, where computing and communication are inseparably intertwined.

\subsection{Adversarial Robustness: From Robust Training to Certified Defense}

\subsubsection{Robustness Against Adversarial Attack}
Adversarial defenses aim to enhance resilience of machine-learning models against malicious perturbations. A significant subclass of these defenses is known as \textit{adversarial immunity}, which focuses on building inherently robust models through targeted interventions during the training or inference phases. This concept can be formally expressed through \textit{astuteness maximization}, where model $f$ is said to be \emph{astute} within the open ball $B_{\eta}(x_0)$ if $f(x') = f(x)$ for all $x' \in B_{\eta}(x_0)$. Higher values of $\eta$ correspond to a stronger robustness against adversarial noise.

However, since directly maximizing astuteness is computationally intensive, it requires the model to be trained over a wide range of perturbations. To alleviate this complexity, various practical strategies have been developed. One such approach is \textit{adversarial training (AT)}, where the training process is guided by the worst-case perturbation derived from a predefined threat model. Rather than optimizing over all possible directions in $B_{\eta}(x_0)$, AT focuses on a representative adversarial example that poses the greatest threat, thus offering a computationally feasible approximation of astuteness maximization.

\subsubsection{Design Recommendations for Unified Security}

To mitigate adversarial threats, a unified security design should integrate both proactive and reactive defenses. The methods discussed below are interlinked and form a comprehensive framework:

\begin{itemize}
    \item \textbf{Robust Training}: Robust training strategies, such as adversarial training \cite{madry2018towards} and Gaussian noise injection \cite{goodfellow2015explaining}, involve exposing the model to perturbations during training so as to enhance generalization and build resilience against a wide array of attacks. Despite its computational intensity, adversarial training remains one of the most effective empirical defense mechanisms.

    \item \textbf{Denoising Autoencoders (DAEs)}: DAEs \cite{vincent2008extracting} learn to reconstruct clean representations from perturbed inputs, thereby reducing adversarial noise. DAEs are particularly useful in systems where inputs have an inherent signal structure, such as images or time-domain waveforms.

    \item \textbf{Certified Defenses}: Certified defenses like randomized smoothing \cite{cohen2019certified} and convex outer polytope bounds \cite{wong2018provable} offer formal guarantees on robustness within bounded perturbation norms. These techniques provide provable security levels, although they typically support only small perturbation sizes.

    \item \textbf{Cross-layer Awareness}: In communication systems, incorporating physical-layer perturbation models into higher-layer ML pipelines leads to more robust end-to-end designs. As demonstrated in \cite{shi2023adversarial}, cross-layer optimization improves resistance to both signal-level and semantic-level adversarial threats.

    \item \textbf{Online Detection and Adaptation}: Real-time adversarial detection approaches, including adversarial drift monitoring \cite{hendrycks2017baseline} and statistical anomaly detection \cite{ma2018characterizing}, enable dynamic model reconfiguration. Overall, these lightweight defenses are particularly suitable for edge devices and resource-constrained environments.

    \item \textbf{Ensemble and Input Transformation Techniques}: Techniques such as feature squeezing \cite{xu2018feature}, JPEG compression \cite{guo2018countering}, and stochastic transformations \cite{dhillon2018stochastic} weaken the effect of adversarial perturbations. Ensemble methods aggregate predictions from diverse models to further dilute attack effectiveness.

    \item \textbf{Input Reconstruction with GAN or AE}: Generative models, such as Defense-GAN \cite{samangouei2018defense} and MagNet \cite{meng2017magnet}, project inputs back onto the data manifold to remove adversarial artifacts. These models hold much promise in semantic communications and image-based classification tasks.
\end{itemize}

The incorporation of domain-specific constraints, such as power constraints and signal structure, is essential to tailoring the design of defenses in wireless systems. These strategies provide a foundation for the development of cross-compatible and future-proof security measures.

\subsubsection{Adversarial Defense in Wireless Communication}

\paragraph{Adversarial defenses on communication in 6G wireless networks}
As 6G networks adopt AI-native protocols, reconfigurable intelligent surfaces, and integrated sensing and communication, they become increasingly vulnerable to adversarial attacks targeting communication reliability and signal integrity. To counteract these threats, previous research explored several adversarial defense strategies operating on different layers of the communication stack. One foundational defense is adversarial training, where models are retrained with adversarial examples to improve robustness. In \cite{bair2022adversarial}, the authors emphasized adversarial training as an effective defense in RF deep learning, particularly for modulation classification and signal authentication tasks. In parallel, model certification methods were proposed to provide formal robustness guarantees under bounded perturbations; however, their applicability to dynamic wireless channels remains limited. To address this concern, in \cite{park2022robust}, the authors proposed robust beamforming algorithms using worst-case adversarial perturbation bounds, thus ensuring resilience in physical-layer attacks. Another line of defense uses signal reconstruction techniques: in \cite{shafiee2021towards}, the authors designed autoencoder-based communication systems capable of detecting and correcting adversarial perturbations before decoding. On the system level, \cite{zhang2023layereddefense} introduced a cross-layer defense framework for 6G, combining physical-layer anomaly detection, secure model aggregation in FL, and protocol adaptation to isolate adversarial influence. Collectively, these studies highlight the importance of multi-level and adaptive defenses in the design of resilient 6G communication systems, particularly considering their reliance on real-time AI inference and distributed computation.

Furthermore, in \cite{bayesiandefense}, the authors constructed a Bayesian neural network (BNN) for AMC, adding a regularization term on the posterior weight variance to prevent extreme weights, as well as to stabilize the model's predictions under adversarial perturbation. The network uses a flexible Sinh–Arcsinh Gaussian prior, which allows control over distribution skewness and tail heaviness and offers a trade-off between robustness and accuracy. Evaluated on standard white-box attacks (FGSM, PGD, Auto‑PGD), the BNN significantly outperforms conventional models, particularly in low perturbation-to-noise ratio scenarios, thus showcasing improved robustness without sacrificing clean accuracy.

\paragraph{Adversarial defenses on sensing in 6G wireless networks}
As 6G networks integrate sensing and communication through ISAC systems and deploy deep learning models for radar perception, localization, and environmental awareness, they face increasing risks from adversarial attacks on sensing components. Defending against these threats requires robust, real-time, and adaptable mechanisms capable of protecting deep learning models against both digital and physical perturbations. In \cite{liu2023adversarial}, the authors proposed adversarially robust radar classification methods by incorporating perturbation constraints into the loss function, thus improving resilience in AI-driven radar sensing tasks. To counter physical-world adversarial examples such as ghost target generation, in another relevant study \cite{chen2023physical}, the authors recommended sensor fusion defenses combining radar with auxiliary modalities (e.g., light detection and ranging (LiDAR) or vision) to detect inconsistencies across modalities. In the domain of automotive ISAC, \cite{hu2022security} introduced a two-stage anomaly detection mechanism that uses temporal sequence consistency to identify abnormal sensing behavior caused by adversarial interference. Furthermore, in \cite{shafiee2021towards}, the authors designed end-to-end robust learning architectures based on autoencoders for radar signal reconstruction and adversarial denoising. Similarly, in \cite{zhang2023layereddefense}, the authors further proposed a layered sensing defense strategy for 6G, including redundancy in spatial observations, model uncertainty quantification, and adaptive retraining using edge-collected adversarial patterns. These strategies highlights that, in order to secure sensing capabilities of 6G networks against sophisticated adversarial attacks, there is a clear need need for combining AI robustness, physical redundancy, and multi-modal fusion .

\paragraph{Adversarial defenses on computing in 6G wireless networks}
The AI-native design of 6G networks incorporates distributed intelligence through on-device inference, FL, and edge-cloud collaboration. This decentralized computing architecture significantly increases the attack surface, particularly from adversarial clients and poisoned updates. To mitigate these threats, several defense mechanisms were proposed. In \cite{li2021federated}, the authors outlined robust aggregation techniques for FL, including Krum and trimmed mean, which limit the influence of malicious clients during global model updates. Furthermore, in \cite{fung2020limitations}, the authors introduced the FoolsGold algorithm that detects colluding adversaries by analyzing similarity in update directions, thereby effectively reducing backdoor threats in non-IID settings. In a 6G context, another pertinent research \cite{zhang2023layereddefense} proposed a layered computing defense strategy combining secure model aggregation, reputation tracking, and edge-based anomaly detection to defend against both inference-time and training-time adversarial manipulation. Moreover, in \cite{bhagoji2019analyzing}, the authors analyzed adversarial robustness of FL under white-box and model-replacement attacks, motivating the use of differential privacy and certified robustness techniques. In their entirety, the methods briefly reviewed above highlight the importance of adversarial resilience in 6G computing systems, where model poisoning, gradient manipulation, and inference attacks may propagate across distributed AI components and compromise the reliability of the entire network.

\subsubsection{Future Research Directions}

The following research gaps must be addressed to advance secure wireless AI systems:

\begin{itemize}
\item \textbf{Channel-aware attacks and defenses}: Future methods should tightly couple physical channel characteristics (e.g., fading, SNR, MIMO properties) with both attack generation and defensive learning.
\item \textbf{Data-efficient defenses}: Meta-learning and few-shot techniques should be used to build adaptive defenses with minimal data overhead.
\item \textbf{Benchmarking and standardization}: There is a clear need for the development of comprehensive evaluation frameworks and public datasets to evaluate and compare adversarial robustness in RF settings.
\item \textbf{Joint optimization with resource constraints}: Further research is needed on low-complexity defenses compatible with real-time hardware, particularly for IoT and edge devices.
\item \textbf{Adversarial training under mobility and channel dynamics}: Adversarial robustness under fast-varying environments should be extended by incorporating dynamic feedback into training loops.
\item \textbf{Hybrid defense architectures}: Model-level robustness (robust training) should be combined with signal-level defenses (autoencoders, spectrum filtering) and environment-level tools (adaptive re-transmission, channel hopping).
\item \textbf{Explainable defenses}: There is a clear need for  the development of explainable AI tools to interpret the effects of perturbations and guide defensive strategies.
\end{itemize}

Future research directions outlined above will bridge the gap between theoretical robustness and real-world deployability. Considering that, adversarial attacks represent a serious security challenge for future ML-augmented wireless networks, a unified design should bridge ML robustness, physical-layer constraints, and communication-specific characteristics. By embracing adaptive, efficient, and cross-layer approaches, researchers can build resilient wireless intelligence systems capable of withstanding adversarial manipulations.

%%%% end 

\section{Conclusion} \label{Sec:conclusion}
6G is anticipated to introduce capabilities that will greatly improve connectivity for everyday users and vertical sectors by supporting immersive and massive communication, HRLLC, ISAC, and AIAC. It is further expected to enable pervasive intelligence, dissolving the traditional boundary between computing and communications. Supporting both training and inference for emerging AI solutions, alongside providing connectivity for intelligent AI agents, will demand a transition away from merely transmitting signals with high fidelity towards semantic and goal-oriented communication that prioritizes downstream task performance. To fully realize these paradigm shifts, it is essential that users can benefit from these new capabilities without compromising security or privacy. In this paper, we first outlined a broad range of potential security threats and privacy issues linked to these technologies. We then offered guidelines on how to achieve secure communication, sensing, and computing by mitigating these risks. In addition, the paper highlighted overarching security- and privacy-aware design principles that apply across all three functionalities and beyond.

Given how integral communication and computing technologies have become to everyday life, and considering the ever-increasing amounts of sensitive information they process, the ongoing struggle between malicious actors and defensive mechanisms is unlikely to conclude in the near future. Adversaries will persist in refining their tactics to bypass new security measures, compelling researchers and practitioners to continuously update and adapt their mitigation approaches in response to emerging threats.

As emphasized throughout this paper, designing a secure next-generation network that offers users ever-greater capabilities alongside enhanced protection raises numerous open challenges and exciting research directions. Addressing most of these issues demands interdisciplinary collaboration that spans multiple areas, including information theory, communications, signal processing, cryptography, and machine learning. Looking forward, the development of quantum-resistant algorithms is an urgent priority as quantum computing and quantum communication continue to evolve. Privacy-preserving, low-complexity computation methods—such as homomorphic encryption and secure multi-party computation—must be further advanced to support meaningful analysis of sensitive data without sacrificing individual privacy. Securing AI-enabled communication networks against adversarial attacks and ensuring algorithmic fairness, explainability, and privacy protection, calls for close cooperation among computer scientists, communication and information theorists, and domain specialists in specific application areas. In a related direction, as biometric authentication and brain–computer interfaces become more widespread, researchers need to confront the distinct privacy challenges of permanent biological identifiers, as well as the associated security vulnerabilities. Lastly, designing security and privacy solutions that are both usable and understandable for non-experts remains a pervasive difficulty across all fields. Pursuing these research avenues will be essential for preserving trust and security in an increasingly interconnected world driven by 6G technologies.

\bibliographystyle{IEEEtran}
\bibliography{Reference}

\begin{IEEEbiography}
[{\includegraphics[width=1in,height=1.25in,clip,keepaspectratio]{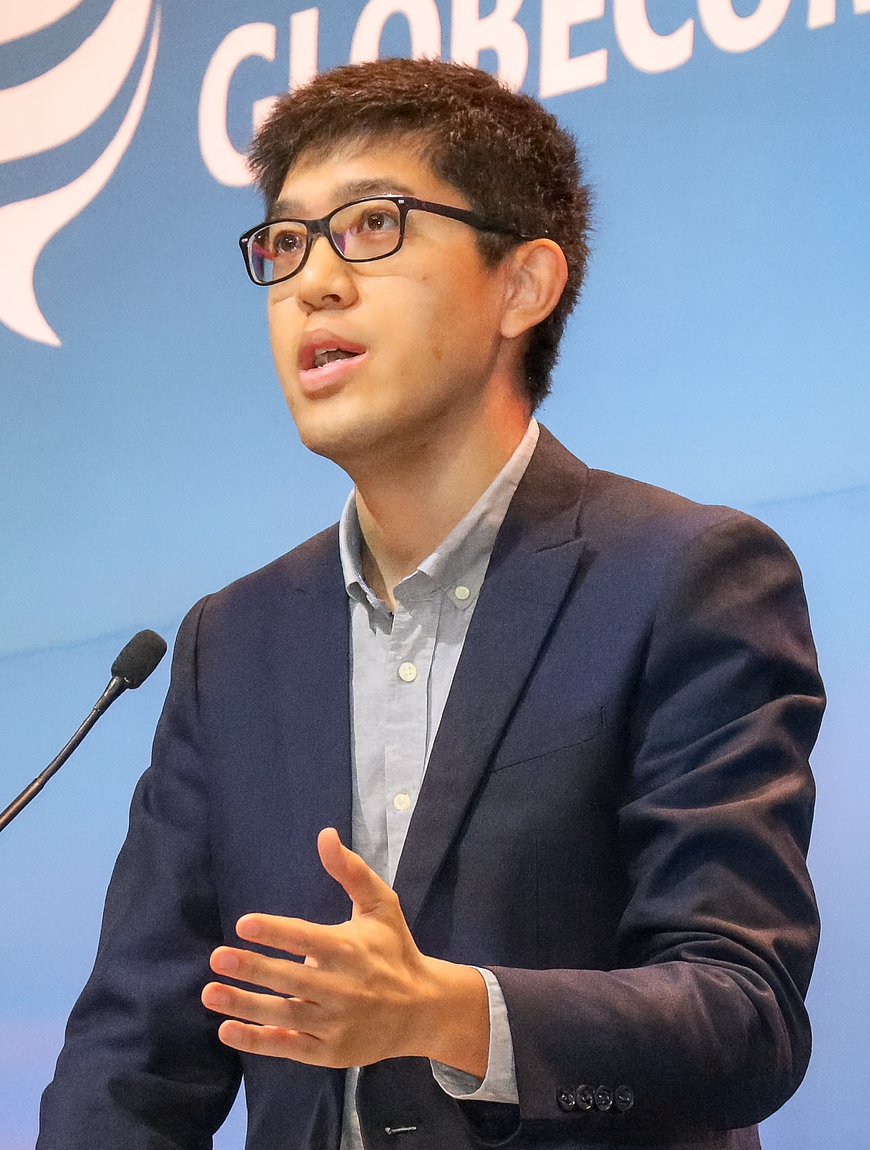}}]{Ruiqi (Richie) Liu} (S'14-M'20-SM'24) is a master researcher in the Wireless and Computing Research Institute of ZTE Corporation. His main research interests include reconfigurable intelligent surfaces, integrated sensing and communication and 6G requirements. He is the author or co-author of several books and book chapters. He has made significant contributions to standardization of 5G / 5G-advanced in 3GPP by authoring and submitting more than 500 technical documents with over 100 approved, and serving as a rapporteur. He served as the chair of multiple correspondence and drafting groups in ITU-R WP5D towards 6G. He currently serves as the Vice Chair of ISG RIS in the ETSI. He is involved in organizing committees of international conferences and is invited to give multiple talks, including the keynote speech at IEEE Globecom 2024. He takes multiple leadership roles in the committees and boards in IEEE ComSoc and VTS, including the voting member of the ComSoc industry communities board. He served as the Deputy Editor-in-Chief of \textsc{IET Quantum Communication}, the Associate Editor for \textsc{IEEE Communications Letters}, the Associate Editor for \textsc{IEEE Communications Magazine} and Guest Editor or Lead Guest Editor for a series of special issues. His recent awards include the 2023 Beijing Science and Technology Invention Award, 2025 IEEE SPCC Early Achievement Award and the Best Paper Award from  \textsc{Intelligent and Converged Networks} (2025). He is listed in the World's Top 2\% Scientists by Stanford/Elsevier.
\end{IEEEbiography}

\begin{IEEEbiography}[{\includegraphics[width=1in,height=1.25in,clip,keepaspectratio]{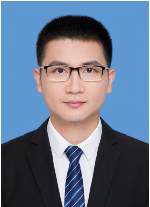}}]{Beixiong Zheng}
(M'18-SM'23) received his B.Eng. and Ph.D. degrees from the South China University of Technology, Guangzhou, China, in 2013 and 2018, respectively. 
He is currently a Professor with the School of Microelectronics, South China University of Technology. He was a Research Fellow with the Department of Electrical and Computer Engineering, National University of Singapore from 2019 to 2022. From 2015 to 2016, he was a Visiting Student Research Collaborator with Columbia University, New York, NY, USA. 
His recent research interests include 5G/6G wireless communications, rotatable antenna (RA), intelligent reflecting surface (IRS), and signal processing.

Dr. Zheng is now serving as the Lead Guest Editor for the \textsc{IEEE Journal on Selected Areas in Communications (JSAC)}, 
and an Editor for the \textsc{IEEE Transactions on Wireless Communications (TWC)} and
the \textsc{IEEE Communications Letters (CL)}.
He was the recipient of
the \textsc{IEEE Communications Society} Asia-Pacific Best Young Researcher Award in 2025,
the \textsc{IEEE Communications Society} Heinrich Hertz Award for Best Communications Letter in 2022,
the \textsc{IEEE Communications Society} Best Tutorial Paper Award in 2023,
the Electronic Information Science and Technology Award (First Prize of Natural Science Award) of Guangdong Province in 2022,
the Science and Technology Award (First Prize of Natural Science Award) of Guangdong Communications Society in 2024,
the Best Ph.D. Thesis Award from the China Education Society of Electronics in 2018,
the Best Demo Award from the IEEE/CIC International Conference on Communications in China (ICCC) in 2025,
the Best Paper Award from the IEEE International Conference on Computing, Networking and Communications (ICNC) in 2016,
the Best Paper Award from the IEEE International Conference on Wireless Communications and Signal Processing (WCSP) in 2024,
and the Best Paper Award from the International Conference on Ubiquitous Communication (Ucom)  in 2023.
He was also awarded the Humboldt Research Fellowship for experienced researchers in 2024.
In addition, he was listed as the Highly Cited Researchers by Clarivate,
the Highly Cited Chinese Researchers by Elsevier, and 
the World’s Top 2\% Scientist by Stanford University and Mendeley Data. 
\end{IEEEbiography}

\begin{IEEEbiography}[{\includegraphics[width=1in,height=1.25in,clip,keepaspectratio]{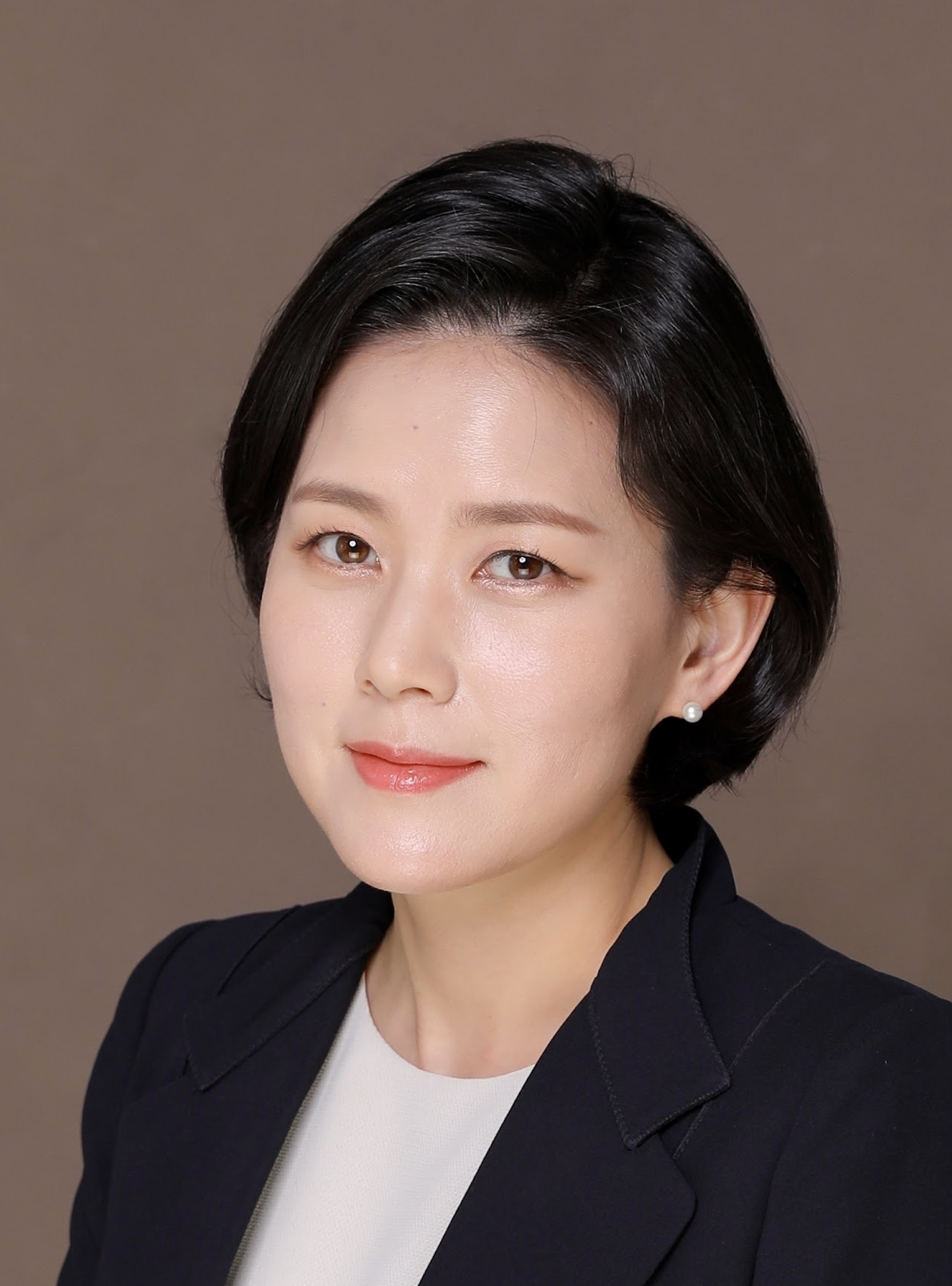}}]
{Jemin Lee} (S`06-M`11-SM'24) received the B.S. (with high honors), M.S., and Ph.D. degrees in Electrical and Electronic Engineering from Yonsei University, Seoul, Korea, in 2004, 2007, and 2010, respectively. 
She was a Postdoctoral Fellow at the Massachusetts Institute of Technology (MIT), Cambridge, MA from 2010 to 2013, a Temasek Research Fellow at iTrust, Centre for Research in Cyber Security, Singapore University of Technology and Design (SUTD), Singapore, from 2014 to 2016. 
She was an Associate Professor at the Department of Electrical Engineering and Computer Science, Daegu Gyeongbuk Institute of Science and Technology (DGIST), Daegu, Korea, from 2016 to 2021, and an Associate Professor at the Department of Electronic and Electrical Engineering, Sungkyunkwan University (SKKU), Suwon, Korea, from 2021 to 2023. 
Currently, she is an Associate Professor at the School of Electrical and Electronic Engineering, Yonsei University, Seoul, Korea. Her current research interests include wireless communications, communication security, intelligent network control, and blockchain technology.  

Dr. Lee is currently an Area Editor for the {\scshape  IEEE Transactions on Machine Learning in Communications and Networking}, and an Editor the {\scshape IEEE Communications Magazine} and the {\scshape IEEE Transactions on Mobile Computing}. She was an Editor for the {\scshape IEEE Transactions on Wireless Communications},  the {\scshape IEEE Transactions on Communications}, the {\scshape IEEE Communication Letters}, and  the {\scshape IEEE Wireless Communications Letters}. She served as a Chair for the IEEE Communication Society (ComSoc) Radio Communications Technical Committee (RCC) from 2021 to 2022. She received the Haedong Young Engineering Researcher Award in 2020, the IEEE ComSoc Young Author Best Paper Award in 2020, the IEEE ComSoc AP Outstanding Paper Award in 2017, the IEEE ComSoc AP Outstanding Young Researcher Award in 2014. 
\end{IEEEbiography}

\begin{IEEEbiography}[{\includegraphics[width=1in,height=1.25in,clip,keepaspectratio]{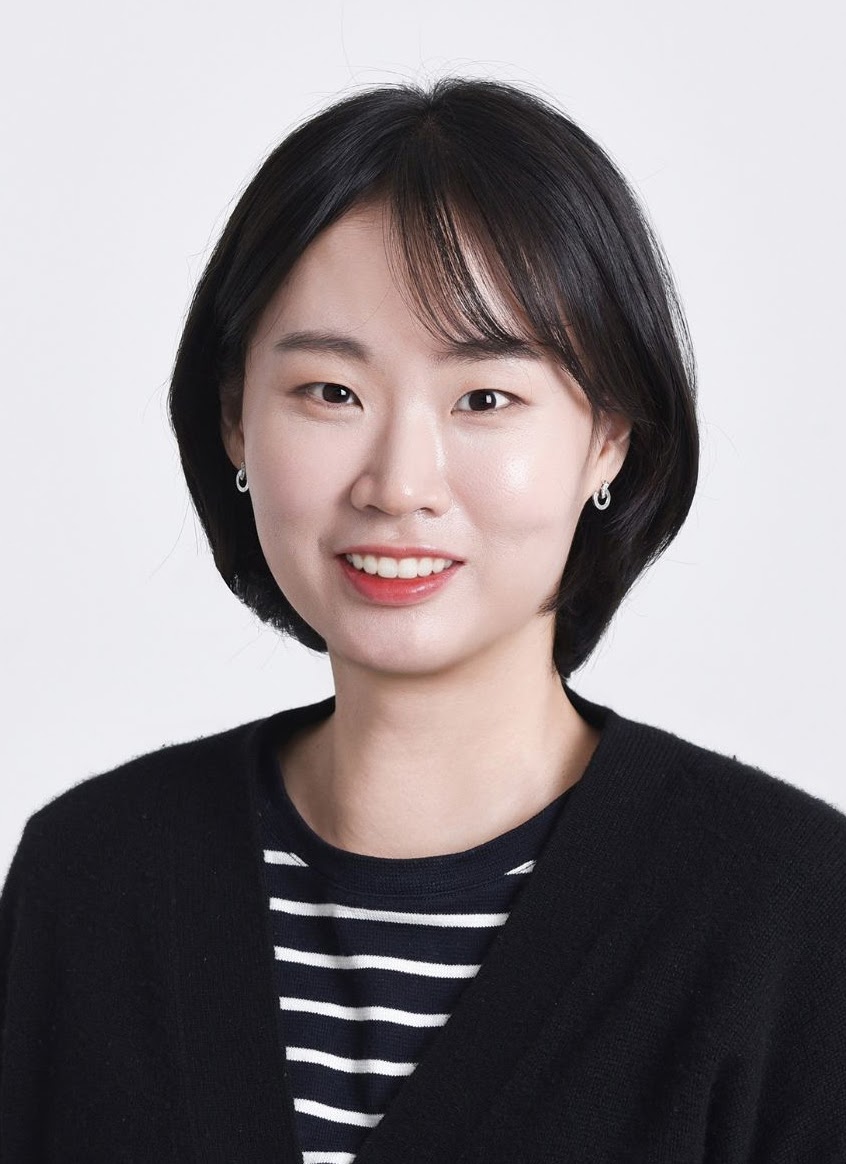}}]
{Si-Hyeon Lee} (S'07-M'13-SM'21) received the B.S. (summa cum laude) and Ph.D. degrees in electrical engineering from the Korea Advanced Institute of Science and Technology (KAIST), Daejeon, South Korea, in 2007 and 2013, respectively. She is currently an Associate Professor with the School of Electrical Engineering, KAIST. She was a Postdoctoral Fellow with the Department of Electrical and Computer Engineering, University of Toronto, Toronto, Canada, from 2014 to 2016, and an Assistant Professor with the Department of Electrical Engineering, Pohang University of Science and Technology (POSTECH), Pohang, South Korea, from 2017 to 2020. Her research interests include information theory, wireless communications, statistical inference, and machine learning. She is currently an Associate Editor for {\scshape IEEE Transactions on Information Theory} and an Associate Editor for {\scshape IEEE Transactions on Communications}, and recently was a Guest Editor for {\scshape IEEE Journal on Selected Areas in Communications}. She was a TPC Co-Chair of IEEE Information Theory Workshop 2024 and an IEEE Information Theory Society Distinguished Lecturer (2024-2025). 
\end{IEEEbiography}

\begin{IEEEbiography}[{\includegraphics[width=1in,height=1.25in,clip,keepaspectratio]{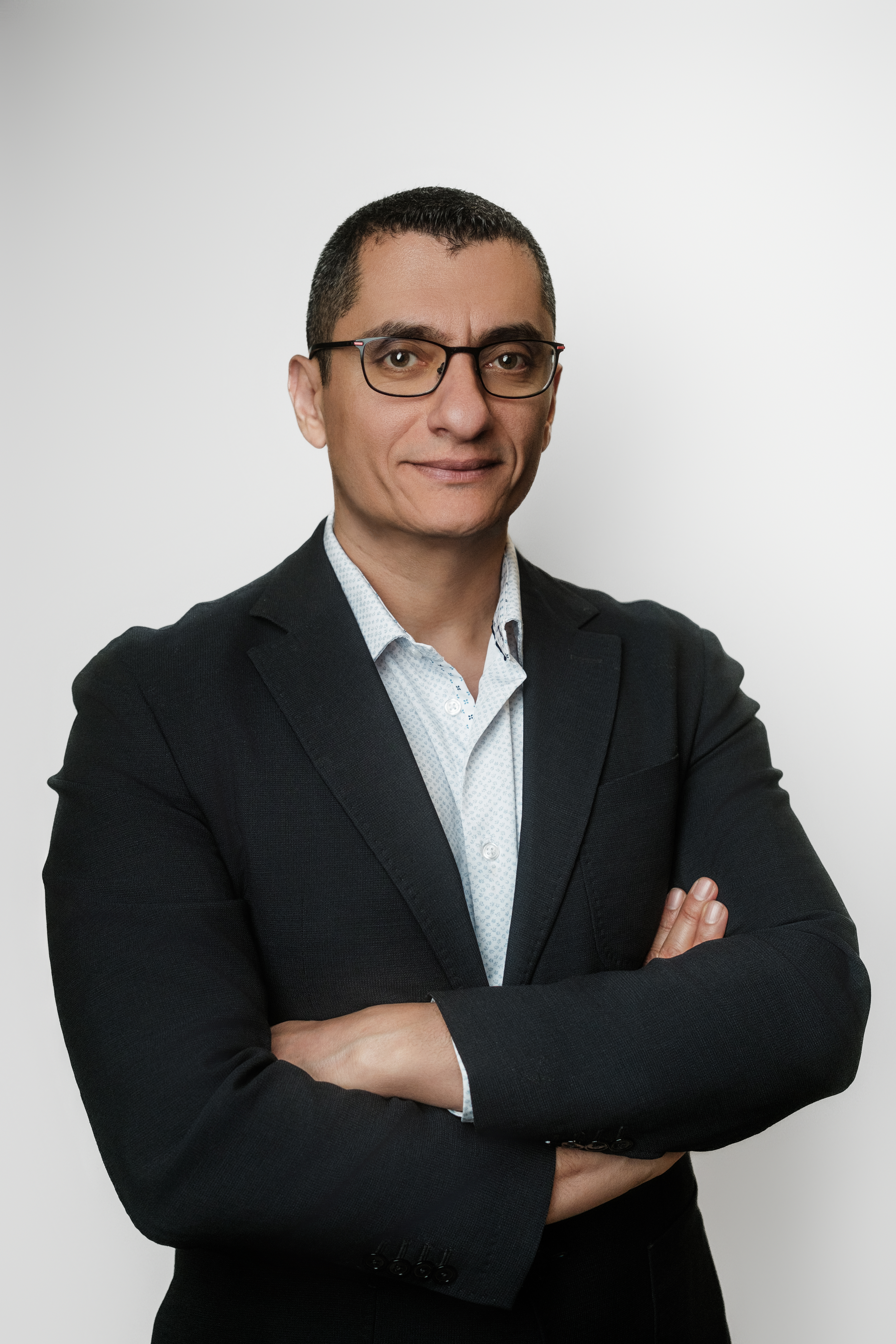}}]
{Georges Kaddoum} (M'09-SM'20) received the bachelor’s degree in electrical engineering from the École Nationale Supérieure de Techniques Avancées, Brest, France, the M.S. degree in telecommunications and signal processing from the Université de Bretagne Occidentale and Télécom Bretagne, Brest, in 2005, and the Ph.D. (with high honors) degree in signal processing and telecommunications from the National Institute of Applied Sciences, University of Toulouse, Toulouse, France, in 2009. He is currently a Professor and the Research Director of the Resilient Machine Learning Institute, holding the Tier 2 Canada Research Chair with the École de Technologie Supérieure (ÉTS), Canada. He has a publication record with over 300 journal articles and conference papers, two chapters in books, and eight pending patents. His recent research focuses on wireless communication networks, tactical communications, resource allocations, and network security. He has received several awards, including Best Paper Awards at prestigious conferences such as IWCMC 2023, PIMRC 2017, and WiMob 2014. Notably, he has been recognized with the IEEE Transactions on Communications Exemplary Reviewer Award in 2019, 2017, and 2015, as well as the Research Excellence Award from the Université du Québec in 2018. In 2019 and 2025, he received the Research Excellence Award from ÉTS. His outstanding contributions were further acknowledged with the 2022 IEEE Technical Committee on Scalable Computing Award for Excellence (Middle Career Researcher), the 2023 MITACS Award for Exceptional Leadership, and the Gold Medal Award from Engineers Canada. He has served as an Associate Editor for the \textsc{IEEE Transactions on Information Forensics and Security} and \textsc{IEEE Communications Letters}. He is currently an Area Editor of the\textsc{ IEEE Transactions on Machine Learning in Communications and Networking}, in addition to his role as an Editor for the \textsc{IEEE Transactions on Communications}.
\end{IEEEbiography}

\begin{IEEEbiography}[{\includegraphics[width=1in,height=1.25in,clip,keepaspectratio]{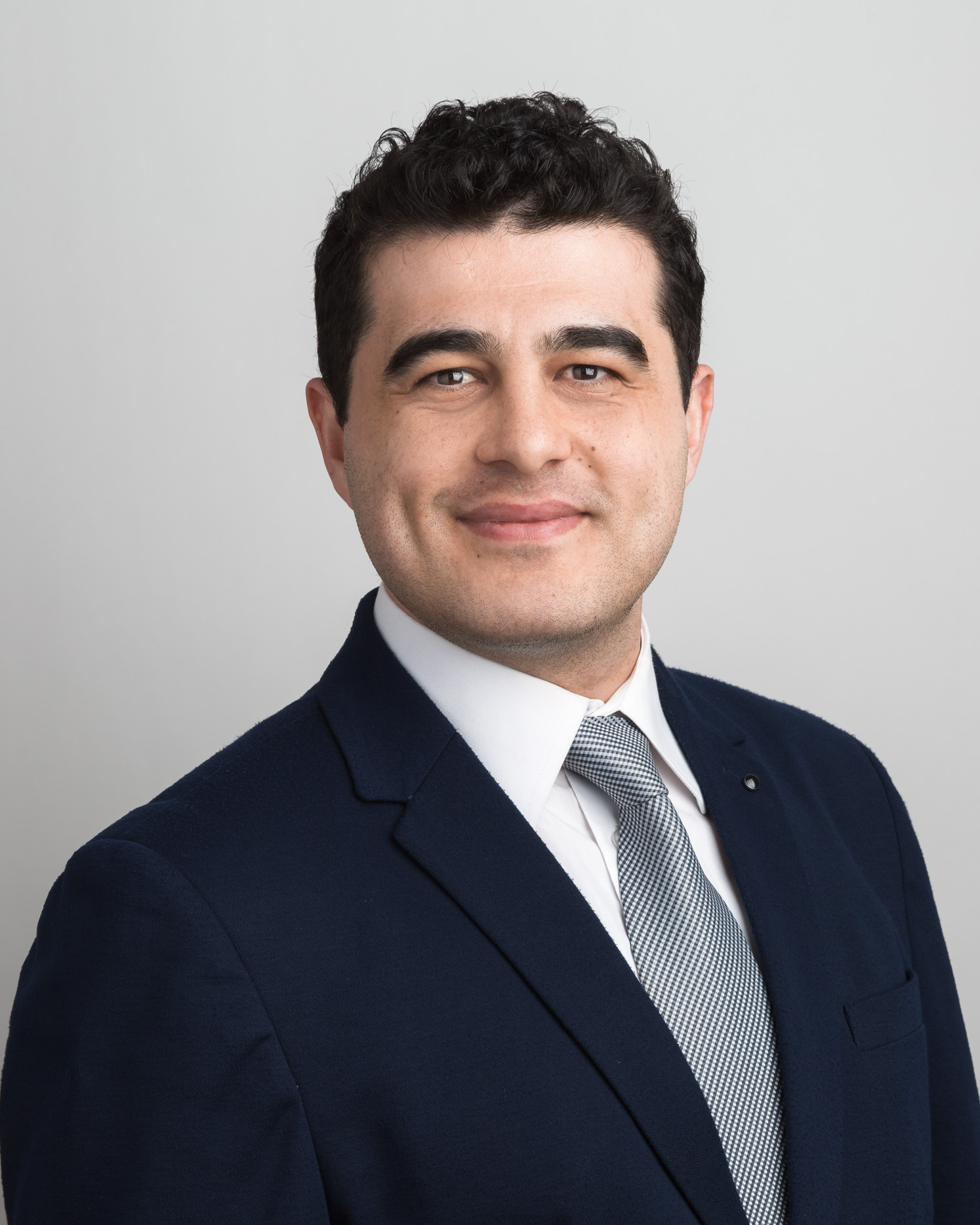}}]{Onur G\"unl\"u} (S'10-M'18-SM'24) received the B.Sc. degree (Highest Distinction) in Electrical and Electronics Engineering from Bilkent University, Turkey in 2011; M.Sc. (Highest Distinction) and Dr.-Ing. (Ph.D. equivalent) degrees in Communications Engineering both from the Technical University of Munich (TUM), Germany in 2013 and 2018, respectively. He was a Working Student in the Communication Systems division of Intel Mobile Communications (IMC), now Apple Inc., in Munich, Germany during November 2012 - March 2013. Onur worked as a Research and Teaching Assistant at TUM Chair of Communications Engineering (LNT) between February 2014 - May 2019. As a Visiting Researcher, among more than twenty Research Stays at Top Universities and Companies, he was at TU Eindhoven, Netherlands during February 2018 - March 2018. Onur was a Visiting Research Group Leader at Georgia Institute of Technology, Atlanta, USA during February 2022 - March 2022. He was also a Visiting Professor at TU Dresden, Germany during February 2023 - March 2023. Following Research Associate and Group Leader positions at TUM, TU Berlin, and the University of Siegen, he joined Linköping University in October 2022 as an ELLIIT Assistant Professor and obtained tenure as an Associate Professor leading the Information Theory and Security Laboratory (ITSL) in August 2024. He obtained the Docent (Habilitation) of Information Theory title in December 2023 and became an IEEE Senior Member in July 2024. Since September 2025, Onur has been a Tenured Full Professor leading the Institute of Communications Engineering (Lehrstuhl für Nachrichtentechnik) at TU Dortmund, Germany. He has received the 2025 IEEE Information Theory Society - Joy Thomas Tutorial Paper Award, the 2023 ZENITH Research and Career Development Award, 2021 IEEE Transactions on Communications - Exemplary Reviewer Award, and the VDE Information Technology Society (ITG) 2021 Johann-Philipp-Reis Award. His research interests include distributed function computation, information-theoretic privacy and security, coding theory, integrated sensing and communication, and private learning. Among his publications is the book \emph{Key Agreement with Physical Unclonable Functions and Biometric Identifiers} (Dr. Hut Verlag, 2019). He serves as an Associate Editor for \textsc{IEEE Journal on Selected Areas in Communications}, \textsc{IEEE Transactions on Communications}, and \textsc{Entropy}, and recently was an Associate Editor for \textsc{EURASIP Journal on Wireless Communications and Networking} and a Guest Editor for \textsc{IEEE Journal on Selected Areas in Information Theory}. He also serves as a Board Member and Secretary of the IEEE Sweden VT/COM/IT Joint Chapter and as a Working Group Leader for EU COST Action 6G Physical Layer Security (6G-PHYSEC).
\end{IEEEbiography}

\begin{IEEEbiography}[{\includegraphics[width=1in,height=1.25in,clip,keepaspectratio]{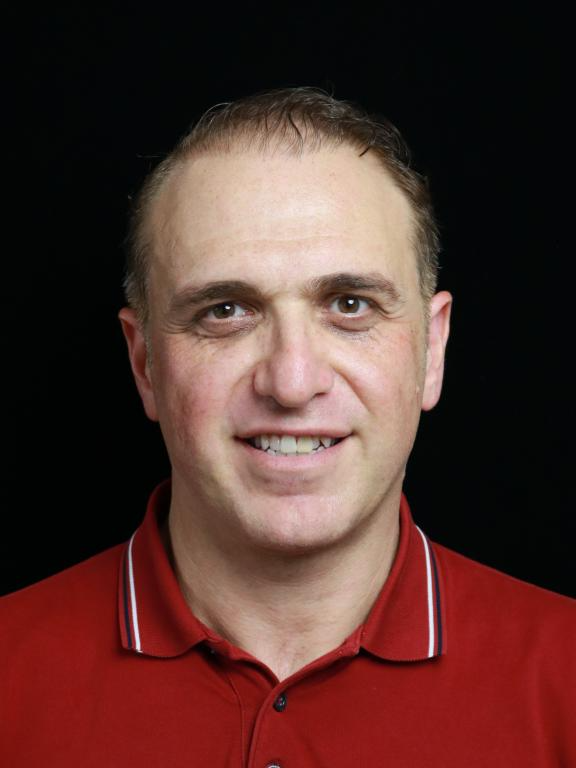}}]{Deniz Gündüz} [S’03-M’08-SM’13-F’22] received the B.S. degree in electrical and electronics 
engineering from METU, Turkey in 2002, and the M.S. and Ph.D. degrees in electrical engineering from NYU Tandon School of Engineering in 2004 and 2007, respectively. In 2012, he joined the Electrical and Electronic Engineering Department at Imperial College London,  UK, where he is currently a Professor of Information Processing, and serves as the deputy 
head of the Intelligent Systems and Networks Group. In the past, he held positions at the  University of Modena and Reggio Emilia (part-time faculty member, 2019-22), University of Padova (visiting professor, 2018, 2020), Centre Tecnologic de Telecomunicacions de Catalunya (CTTC) (research associate, 2009-12), Princeton University (postdoctoral researcher, 2007-09, visiting researcher, 2009-11) and Stanford University (research assistant professor, 2007-09). His research interests lie in the areas of information theory, machine learning, wireless communications and privacy. Dr. Gündüz is a Fellow of the IEEE. He is an elected member of the IEEE Signal Processing Society Signal Processing for communications and Networking (SPCOM) and Machine Learning for Signal Processing (MLSP) Technical Committees. He 
chairs the UK and Ireland Chapter of the IEEE Information Theory Society, and serves as an area editor for the IEEE Transactions on Information Theory. In the past, he served in 
editorial roles for the IEEE Transactions on Communications, IEEE Transactions on Wireless Communications, IEEE Journal on Selected Areas in Communications and IEEE Journal on 
Selected Areas in Information Theory. He is the recipient of the IEEE Communications Society - Communication Theory Technical Committee (CTTC) Early Achievement Award in 2017, Starting (2016), Consolidator (2022) and Proof-of-Concept (2023) Grants of the European Research Council (ERC), and has co-authored several award-winning papers, most recently the IEEE Communications Society - Young Author Best Paper Award (2022), and the IEEE International Conference on Communications Best Paper Award (2023). He received the 
Imperial College London - President's Award for Excellence in Research Supervision in 2023.
\end{IEEEbiography}

\end{document}